\documentclass[sigplan,screen,natbib=false,9pt]{acmart}
\usepackage{cite}
\usepackage{amsmath,amsfonts}
\usepackage{algorithm,algorithmic}
\usepackage{graphicx}
\usepackage{textcomp}
\usepackage{xcolor}

\usepackage{subcaption}

\usepackage{tablefootnote}
\usepackage{url}

\usepackage{enumitem}
\setitemize{noitemsep,topsep=0pt,parsep=0pt,partopsep=0pt}

\usepackage{array}
\newcolumntype{P}[1]{>{\centering\arraybackslash}p{#1}}
\newcolumntype{M}[1]{>{\centering\arraybackslash}m{#1}}

\usepackage[numbers]{natbib}

\renewcommand{\paragraph}[1]{\noindent\vspace{0.2em}\textbf{\emph{#1}}}

\copyrightyear{2022}
\acmYear{2022}
\setcopyright{rightsretained}
\acmConference[Middleware '22]{23rd International Middleware
Conference}{November 7--11, 2022}{Quebec, QC, Canada}
\acmBooktitle{23rd International Middleware Conference (Middleware'22), November 7--11, 2022, Quebec, QC, Canada}
\acmDOI{10.1145/3528535.3565248}
\acmISBN{978-1-4503-9340-9/22/11}

\begin{document}

\title{CGX: Adaptive System Support for Communication-Efficient Deep Learning}

\author{Ilia Markov}
\email{ilia.markov@ist.ac.at}
\affiliation{
    \institution{Institute of Science and Technology Austria}
    \city{Klosterneuburg}
    \country{Austria}
}
\author{Hamidreza Ramezanikebrya}
\authornote{Work performed  during an internship at Institute of Science and Technology Austria.}
\email{hamid@ece.ubc.ca}
\affiliation{
    \institution{University of British Columbia}
    \city{Vancouver}
    \country{Canada}
}

% \author{Hamidreza Ramezanikebrya}
% \email{hamid@ece.ubc.ca}
% \affiliation{
%     \institution{Institute of Science and Technology Austria}
%     \city{Klosterneuburg}
%     \country{Austria}
% }

\author{Dan Alistarh}
\email{dan.alistarh@ist.ac.at}
\affiliation{
    \institution{Institute of Science and Technology Austria}
    \city{Klosterneuburg}
    \country{Austria}
}

\renewcommand{\shortauthors}{Markov et al.}

\begin{abstract}
The ability to scale out training workloads has been one of the key performance enablers of deep learning. The main scaling approach is data-parallel GPU-based training, which has been boosted by hardware and software support for highly efficient point-to-point communication, and in particular via hardware bandwidth overprovisioning. 
Overprovisioning comes at a cost: there is an order of magnitude price difference between ``cloud-grade'' servers with such support, relative to their popular ``consumer-grade'' counterparts, although single server-grade and consumer-grade GPUs can have similar computational envelopes. 

In this paper, we show that the costly hardware overprovisioning approach can be supplanted via algorithmic and system design, and propose a framework called CGX, which provides efficient software support for compressed communication in ML applications, for both multi-GPU single-node training, as well as larger-scale multi-node training. 
CGX is based on two technical advances: 
\emph{At the system level}, it relies on a re-developed communication stack for ML frameworks, which provides flexible, highly-efficient support for compressed communication. \emph{At the application level}, it provides \emph{seamless, parameter-free} integration with popular frameworks, so that end-users do not have to modify training recipes, nor significant training code. 
This is complemented by a \emph{layer-wise adaptive compression} technique which dynamically balances compression gains with accuracy preservation. 
CGX integrates with popular ML frameworks, providing up to 3X speedups for multi-GPU nodes based on commodity hardware, and order-of-magnitude improvements in the multi-node setting, with negligible impact on accuracy.

% We show that this framework is able to remove communication bottlenecks from data-parallel DNN training, in the absence of hardware support: when training modern models and tasks to full accuracy, CGX provides self-speedups of 2-3X for an 8-GPU commodity node, enabling commodity hardware to surpass the throughput of a much more expensive NVIDIA DGX-1 server. In the multi-node setting, CGX enables order-of-magnitude speedups by identifying and solving a novel \emph{layer-wise adaptive compression problem}, in which we can automatically set compression levels in a layer-wise fashion, balancing speedup and accuracy recovery. 
 \end{abstract}

\begin{CCSXML}
<ccs2012>
   <concept>
       <concept_id>10010147.10010919.10010172</concept_id>
       <concept_desc>Computing methodologies~Distributed algorithms</concept_desc>
       <concept_significance>500</concept_significance>
       </concept>
   <concept>
       <concept_id>10010147.10010257.10010293.10010294</concept_id>
       <concept_desc>Computing methodologies~Neural networks</concept_desc>
       <concept_significance>300</concept_significance>
       </concept>
 </ccs2012>
\end{CCSXML}

\ccsdesc[500]{Computing methodologies~Distributed algorithms}
\ccsdesc[300]{Computing methodologies~Neural networks}
\keywords{Distributed Systems, Deep Learning, Gradients compression}

\maketitle

%TODO remove page numbers
% \thispagestyle{plain}
% \pagestyle{plain}

\section{Introduction}

Deep learning has made significant leaps in terms of accuracy and performance, enabled by the ability to scale out workloads. Yet, distributed scalability of deep neural network (DNN) training still presents non-trivial challenges, and the last decade has seen a tremendous amount of work on distributed paradigms, algorithms, and implementations to address them~\cite{ li2014scaling, chilimbi2014project, abadi2016tensorflow, peng2019generic, jiang2020unified}.
% Training a neural network is usually split into a \emph{forward pass}, generating output predictions over some data samples, and a \emph{backward pass}, producing gradient updates based on the ground truth. Arguably, the standard distribution strategy is \emph{data-parallel}, in which individual nodes compute gradients on data samples in parallel, and then aggregate gradients. 
Specifically, two key scaling challenges behind are reducing the \emph{synchronization costs} among computing nodes~\cite{jayarajan2019priority, jiang2020unified, peng2019generic, li2021sync}, and minimizing the \emph{communication costs} which arise naturally due to the high bandwidth requirements of all-to-all transmission of model updates (gradients) between nodes.
In this paper, we focus mainly on mitigating the \emph{bandwidth cost} of gradient transmission in DNN training, which is an increasingly common bottleneck, correlated to the soaring parameter counts of modern machine learning models.

% Both problems have received significant research attention. A standard approach for reducing synchronization is fine-grained scheduling, for which high-performance solutions are known~\cite{jayarajan2019priority, jiang2020unified, peng2019generic, li2021sync}. 
% Yet, this approach does not mitigate the  case where the \emph{bandwidth cost} of gradient transmission is dominant. 
% We focus on this increasingly common bottleneck in this work, which is correlated to the soaring parameter counts of modern machine learning models. 

There are two main strategies for removing bandwidth bottlenecks. 
The \emph{industrial approach} has been to employ \emph{bandwidth over-provisioning}: for instance, the inter-GPU bandwidth for NVIDIA-enabled cloud-grade multi-GPU servers has increased by more than 30X between 2015 (Kepler generation) and the post-2018 Ampere generation, and has been complemented by a customized GPU-centric communication library, called NCCL, which leverages hardware support. 
%Competing hardware alternatives in the cloud, such as TPUs or IPUs, have taken similar approaches. 
Yet, bandwidth over-provisioning comes at significant hardware and development costs, reflected in the monetary cost borne by end-users: there is an almost order-of-magnitude cost difference between \emph{cloud-grade},  overprovisioned multi-GPU servers such as NVIDIA DGX systems~\cite{DGX} and \emph{commodity} workstations, built using consumer-grade GPUs (e.g. NVIDIA GeForce/RTX series). 
The latter have become extremely popular, due to lower costs and comparable single-GPU performance~\cite{LambdaCloud, LeaderGPU, GenesisCloud}; however, as we show, there are major performance gaps between the two in terms of  scalability. 

% Due to this trend, the elegant algorithmic  solutions mentioned above have so far largely failed to gain traction in practice: to our knowledge, only one such approach (gradient decomposition via PowerSGD~\cite{vogels2019powersgd}) is supported natively by one popular framework (PyTorch). 

The alternative \emph{algorithmic approach} builds on the fact that stochastic gradient descent (SGD), the standard algorithm for neural network training, can converge with \emph{compressed} gradients. Several elegant lossy compression methods, such as gradient quantization~\cite{seide20141, alistarh2017qsgd, wen2017terngrad}, sparsification~\cite{strom2015scalable, dryden2016communication}, and gradient decomposition~\cite{vogels2019powersgd, wang2018atomo}, allow the theoretical bandwidth cost to be reduced by \emph{up to two orders of magnitude} without accuracy loss. 
Despite their promise, realising these gains in practice runs into a number of significant challenges. 

The first challenge is that of \textbf{parametrization and integration}: approaches such as gradient sparsification or decomposition often require non-trivial parameter and implementation changes to the training process, e.g.~\cite{lin2017deep, renggli2019sparcml, vogels2019powersgd}, to support compression. 
This would require practitioners to revisit their entire training setup, and tune additional hyper-parameters, in order to achieve compression while recovering accuracy. 
A second challenge is that of \textbf{efficient system support} for communication-compression, as it often requires significant changes to lower levels of the software stack, such as supporting compressed or sparse data types. 
Despite research in this direction~\cite{grace, bai2021hipress, gan2021bagua}, the question of general and efficient system support for communication-compression is still open: currently, only one such approach, PowerSGD  decomposition~\cite{vogels2019powersgd}, is supported natively by one popular framework, PyTorch~\cite{paszke2019pytorch}.

\paragraph{Contributions.} In this paper, we introduce a communication framework called CGX, which addresses these challenges, and allows for \emph{parameter-free, seamless integration} of communication-compression into data-parallel DNN training workflows, with up to order-of-magnitude speedups for data-parallel DNN training.  

At the application level, CGX starts from an investigation of the feasibility of \emph{parameter-free compression}: specifically, we implement and test all existing algorithmic approaches, and identify a variant of quantization-based compression that converges to \emph{full accuracy} for many popular models, under \emph{fixed, universal settings of parameters}, without modifying to the original training recipes. 
At the system level, we investigate how gradient compression can be seamlessly and efficiently integrated with modern ML frameworks. Specifically, we  revisit the entire communication stack of modern ML frameworks with compression in mind, from a new  point-to-point communication mechanism which supports compressed types, to compression-aware reductions, and finally a communication engine which interfaces with ML frameworks,  supporting compression  at the tensor/layer level. 

The existence of a parameter-free compression technique which recovers accuracy, combined with the ability of CGX to customize the compression level per layer motivates a new \emph{layer-wise adaptive compression problem}. The idea is that we can to customize the way model gradients are compressed in \emph{layer-wise} fashion, so that the overall compression error is close to a given accurate baseline, but maximizing the bandwidth gains: for instance, one can apply more aggressive compression to layers that are larger, but less ``sensitive'' in terms of accuracy. 
While prior work has already considered techniques which adapt the degree of compression during training, e.g.~\cite{Agarwal2021, Abdelmoniem2021}, this is the first instance of this problem to jointly considers both error and compression constraints at the fine-grained \emph{per-layer} level. Our experimental results show that our layer-wise adaptive compression can bring significant additional gains. 

To justify our design choices, we contrast our design against the first implementation of quantized collectives in NCCL, which we call QNCCL, which we contribute as a separate artefact, showing clear performance and usability improvements in favor of the CGX design. In addition, CGX does not require significant user-code or training pipeline changes, as we provide turn-key integrations with popular ML frameworks such as Pytorch and Tensorflow.

% We perform an in-depth analysis of efficient support for algorithmic communication-reduction methods in machine learning workloads. 
% Our study is focused on the standard \emph{data-parallel} approach to distributed DNN training. 
% On the systems side, our main contribution is a new communication framework called CGX, which can remove the bandwidth bottleneck from commodity GPU servers via efficient and accurate support for communication-compression. 
% On the conceptual side, we identify and solve the \emph{adaptive compression problem}, i.e. the problem of identifying layer-wise compression ratios, which maximize the speedup due to compression, while minimizing the accuracy loss during training.   
% Thus, we show that bandwidth over-provisioning is \emph{not necessary} for scaling modern machine learning workloads, and that bottlenecks can be removed algorithmically in a \emph{generic and parameter-free} manner. 

\begin{figure}[t]
     \centering
     \includegraphics[width=0.5\textwidth]{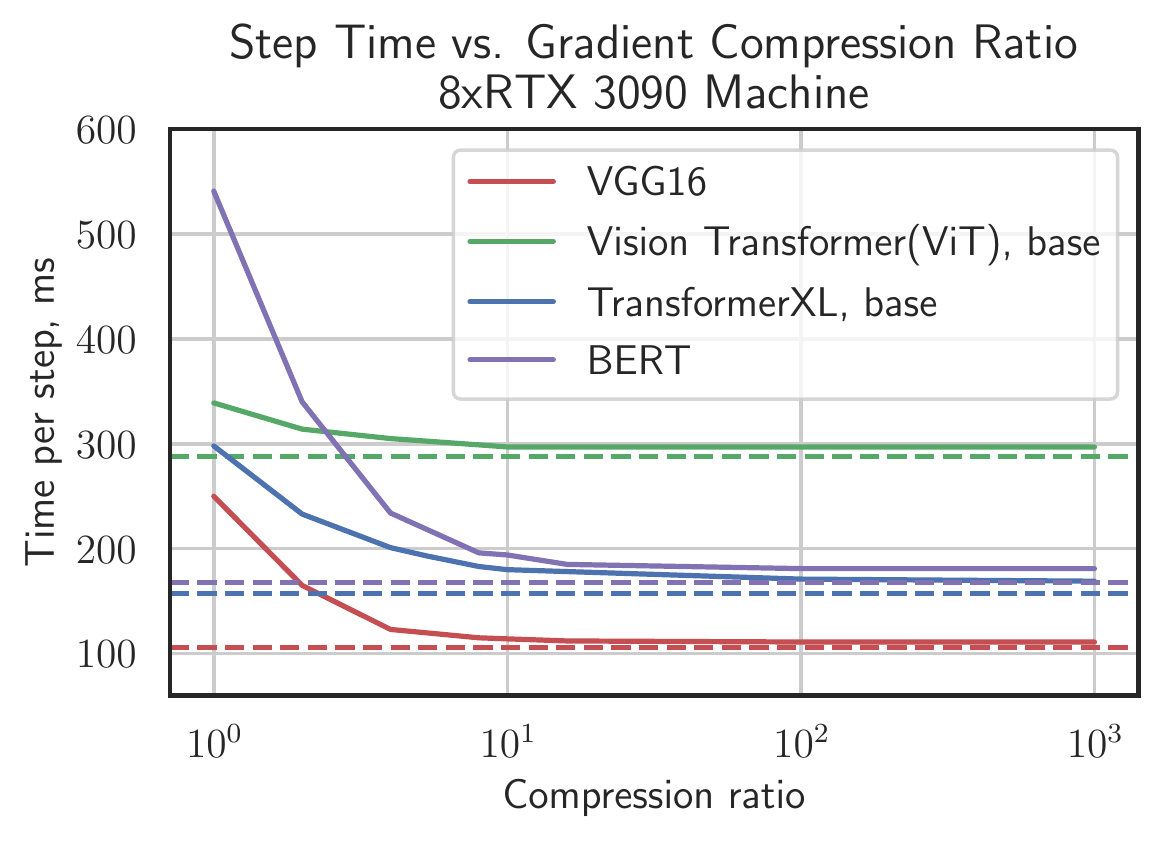}
     \caption{{\small Compression vs. average step time for different models, when using all GPUs on an 8x RTX-3090 machine (Table~\ref{table:systems_char}). Dotted lines denote the throughput at perfect scalability for each model.  Throughput nears  ideal as we decrease transmission size, suggesting that bandwidth is the main bottleneck. See Section~\ref{sec:motivation} for  details.}}
     \label{fig:fake_compression}
 \end{figure}

% We contrast bandwidth-overprovisioned \emph{cloud-grade servers}, with \emph{commodity servers} using consumer-grade GPUs, augmented with efficient algorithmic support for communication-reduction methods. 
\paragraph{Experimental Validation.} 
From the practical perspective, our work is motivated by the experimental data in Figure~\ref{fig:fake_compression}, showing that \emph{bandwidth congestion} is the key scalability bottleneck on single-node, multi-GPU commodity servers, which have emerged as a popular training approach~\cite{LambdaCloud, GenesisCloud, LeaderGPU}. 
The same phenomenon occurs generally in multi-node data-parallel training settings, for a wide range of current and emerging training workloads, from image classification using classical convolutional neural networks (CNNs), to  Transformer-based models for both language modeling~\cite{vaswani2017attention, dai2019transformer} and image classification~\cite{parmar2018image, dosovitskiy2020image}.

%  We begin by examining the scalability bottlenecks  in current and emerging ML training workloads on both types of machines. 
%  Specifically, we examine not just classical convolutional neural networks (ResNets~\cite{he2016deep} and VGG~\cite{simonyan2014very}) on image classification tasks, but also recent Transformer-based models for both language modeling~\cite{vaswani2017attention, dai2019transformer} and image classification tasks~\cite{parmar2018image, dosovitskiy2020image, chen2021vision} as well as models from the BERT family on language modeling tasks. 
 
%  Unsurprisingly, we find that \emph{bandwidth congestion}, is the key scalability bottleneck on commodity servers, whereas it has minor impact for cloud-grade servers: 
%  Figure~\ref{fig:fake_compression} illustrates this issue in the context of the commodity workstation based on 8x 3090 RTX GPUs,  specified in Table~\ref{table:systems_char}. 
 
  We validate our system experimentally in both single-node and multi-node settings, across all of the above standard training tasks. We compare servers using commodity NVIDIA GPUs (RTX series) against cloud-grade NVIDIA servers from the Volta and Ampere architectures. (See Table 2 for details.) 
  First, we find that, once communication bottlenecks are eliminated from ``commodity" machines using CGX, they can match or  \emph{outperform} cloud-grade server with similar peak performance. Importantly, this can be done with negligible accuracy loss. 
 
  For example, we find that, on a commodity 8x RTX 3090 server, CGX can almost \emph{triple} training throughput, reaching up to 90\% of the ideal scaling,  matching or even outperforming a bandwidth-overprovisioned (and more expensive) DGX-1 system. Our second application is to \emph{multi-node training}, where we show up to 10x performance gains, enabled in part by our new solution to the adaptive layer-wise compression problem, without accuracy loss or additional parameters.  
  
  Our findings imply that hardware bandwidth overprovisioning may not be required for scalability in DNN training, and that highly-customized, hyperparameter-heavy compression techniques are not always necessary to remove bottlenecks. This should be immediately useful to users aiming to scale such workloads on commodity or multi-node hardware, but also more broadly for hardware/software co-design for distributed deep learning.

%   In addition, on an 8x A6000 server, CGX successfully supplants NVLINK hardware support, without performance or accuracy loss. 
  
%  Concretely, our system is able to maintain accuracy with average bandwidth compression of 8-10X, and we obtain self-speedups of $2 \times$, and scaling of  $85\%$ relative to perfect scaling, without any hardware support. 

%  Our findings so far are mainly aimed at enabling better scalability on research-dedicated workstations. (This is due to the NVIDIA EULA, which prevents the usage of commodity GPUs in a cloud environment, except for crypto-coin mining.) To address this limitation, we ported our framework to enable multi-GPU scaling on AMD GPUs as well. We find that TBD.

\begin{table*}[t]
\centering
\caption{Server-grade (first 2) vs. consumer-grade NVIDIA GPUs. Throughput  obtained using the NVIDIA Deep Learning Examples benchmark~\cite{DeepLearningExamples}. TDP stands for Thermal design power.}
\label{table:GPU_char}
{\footnotesize
\centering
\begin{tabular}{|c|c|c|c|c|c|c|c|c|}
\hline
 	GPU type    & Arch.  & SM  & TensorCores & GPU Direct & GPU RAM, GB& TDP & ResNet50  & Transformer-XL\\
% \hline
% A100        & Ampere  & 108 & 432 & 3 & 40 & 2470 imgs./s & 60K tokens/s \\
\hline
V100        & Volta   & 80  & 640 & Yes & 16 & 250 Watt & 1226 imgs./s & 37K tokens/s \\
\hline
A6000       & Ampere  & 84  & 336 & Yes & 48 & 3000 Watt & 566 imgs./s & 39K tokens/s \\ 
\hline
RTX 3090    & Ampere  & 82  & 328 & No & 24 & 350 Watt & 850  imgs./s & 39K tokens/s \\
\hline
RTX 2080 TI & Turing  & 68  & 544 & No & 10 & 250 Watt & 484 imgs./s & 13K tokens/s \\
\hline
\end{tabular}
}
\end{table*}

\begin{table*}[t]
    % \begin{tabular}{|l|l|l|l|l|l|l|}
    \centering
    \caption{Systems characteristics of workstations used in evaluation.}
    \label{table:systems_char}
    %\vskip 0.15in
    {\footnotesize
    \centering
\begin{tabular}{|P{1.8cm}|P{1.8cm}|P{2.2cm}|P{3cm}|P{1.7cm}|P{1.4cm}|P{0.8cm}|}
        \hline 
        System & GPUs & Inter-GPU link & Inter-GPU bandwidth & GPU RAM & RAM & CPUs  \\
        \hline
        DGX-1 & 8xV100 & NVLink & 100 GBps & 128 GB & 512 GB & 64 \\
        \hline
        A6000  & 8xA6000 & NVLink & 100 GBps & 384 GB & 1008 GB & 128 \\ 
        \hline
        % p4d.24xlarge & A100 & NVLink & 600 & 320 & 1152 & 96 \\
        % \hline
        RTX-3090  & 8xRTX3090 & None (bus) & 15 GBps & 192 GB & 512 GB & 128 \\
        \hline
        RTX-2080  & 8xRTX2080 TI & None (bus) & 15 GBps & 96 GB & 256 GB & 72 \\

        \hline

    \end{tabular}
    }
\end{table*}

\section{Motivation and Prior Work}
%-------------------------------------------------------------------------------

\subsection{A Motivating Experiment}
\label{sec:motivation}

The standard computational unit for DNN training is the multi-GPU node, usually in instances with  4--16 GPUs. 
End-users often rely on consumer-grade GPUs for training, whereas traditionally cloud services mainly employ cloud-grade GPUs, with some notable exceptions, e.g.~\cite{LambdaCloud, LeaderGPU, GenesisCloud}. 
We begin by briefly examining the scalability differences between cloud and commodity GPU servers. 
As we illustrate in Figures~\ref{fig:throughput-txl} and~\ref{fig:throughput-bert}, the maximum effective throughput of a cloud-grade 8-GPU DGX-1 server is $> 2\times$ higher than that of a comparable commodity 8xRTX-3090 GPU server, when using the same state-of-the-art software configuration (specifically, {Horovod~\cite{sergeev2018horovod} on top of the NCCL communication library}).

This gap is surprising, considering that the single-GPU performance is similar (see Table~\ref{table:GPU_char}). 
To examine the specific impact of \emph{gradient transmission / bandwidth cost}, we implemented a synthetic benchmark that reduces bandwidth cost by artificially compressing transmission. 
Specifically, assuming a buffer of size $N$ to be transmitted, e.g. a layer's gradient, and a target compression ratio $\gamma \geq 1$, we only transmit the first $k = N / \gamma$ elements. 
The results for the 8x RTX-3090 machine, using all 8 GPUs, are shown in Figure~\ref{fig:fake_compression}, where the compression ratio is varied on the X axis, and we examine its impact on the time to complete an optimization step, shown on the Y axis. 
The dotted line represents the time per step in the case of ideal (linear) scaling of single-GPU times. 
We consider Transformer~\cite{dai2019transformer} and BERT-based models~\cite{devlin2018bert} for language modelling tasks, as well as  VGG-16~\cite{simonyan2014very} and Vision Transformer (ViT) models for  classification on ImageNet. 

We therefore observe that \emph{bandwidth cost appears to be the main scalability bottleneck on this machine}. Moreover, recent models (Transformer-XL and ViT) benefit more from compression relative to the classic ResNet50 model, which has  fewer parameters.  Second, \emph{there are limits to how much compression is required for scalability}, which depend on the model characteristics. An order of magnitude compression appears to be sufficient for significant timing improvements, although Transformer-based architectures can still benefit from compression of up to two orders of magnitude.  

\paragraph{Discussion.}
\label{sec:nvlink}
The reason for this poor scalability is the lack of efficient communication support. Specifically,  GPU-to-GPU  transmissions on commodity hardware have significantly lower bandwidth, and higher latency, relative to their cloud counterparts. Specifically, in software , the NVIDIA GPUDirect technology should allow GPUs on the same machine to communicate directly, without the need for extra memory copies. Commodity GPUs, such as the RTX 3090, do not support this technology. 
At the same time, the hardware communication support for NVIDIA GPUs, i.e. NVLink and NVSwitch components, is also not available or severely restricted for commodity GPUs~\cite{rtx3090_nvlink,nvidia_ampere}.
\newpage

% Next, we explore ways of realizing these ideal speedups in practice by allowing for efficient gradient compression while still maintaining the training accuracy of the corresponding models. We begin by providing some background for DNN training and communication compression. 

% \paragraph{Required compression ratio}
% Before picking the compression algorithm and choosing the compression parameters we need to understand what level of compression we need to achieve near linear scalability on our machine. To measure the impact of compression on the training performance we embedded compressor that basically selects the first $k = N \times \gamma$ elements of a buffer, where $N$ size of the buffer, $\gamma$ compression ratio. Figure~\ref{fig:fake_compression} shows that even small compression gives significant speedup.

\subsection{Data-Parallel DNN Training} 

\paragraph{Distribution Strategies and Costs.} Training a DNN essentially minimizes a loss function, related to the error of the model on the dataset, via a sequence of optimization steps, each acting on some data samples.  
% Given a $d$-dimensional model $x_t$, this is performed via a sequence of optimization steps of the form 
% $x_{t + 1} = x_t - \eta_t \nabla f_i (x_t),$  
%  where $\eta_t$ is the learning rate at step $t$ and $\nabla f_i (x_t)$ is the gradient of the loss function with respect to the weights $x_t$ at a specific sample $i$. 
% % Each such optimization step is composed of a \emph{forward pass}, during which the predictions are obtained from the current version of the model $x_t$. The difference between the predictions and the ground truth is measured via the loss function, and an update to the model is computed in the \emph{backward pass}. The gradients are then applied, to obtain the updated model $x_{t + 1}$. 
To preserve computational efficiency, it is common to perform a \emph{batched} version of this process, by which several samples are processed in a single optimization step, and the sum of gradients is applied. 

Data-parallelism is arguably the standard way to scale DNN training, and can be viewed as a variant of batch SGD in which sample gradients are generated in parallel over compute nodes. Specifically, the dataset is partitioned over nodes, each of which maintains a copy of the model, and computes gradients over samples in parallel. Periodically, these gradients are aggregated (e.g., averaged) and the resulting update is applied to all local models. 

Several techniques have been proposed to address the synchronization and communication costs inherent to this lock-step averaging procedure. 
% While our focus is on the communication cost, we first briefly overview methods to reduce synchronization overheads. 
% On one side, asynchronous methods, e.g.~\cite{niu2011hogwild}, allow nodes to completely break lock-step synchronization and proceed roughly independently, at the cost of additional tuning for recovering full accuracy. On the other, synchronous scheduling~\cite{peng2019generic, jiang2020unified} preserves lock-step progress but performs fine-grained scheduling of communication and computation, sometimes at the sub-layer level, maximizing the overlap between computation and communication. 
Here, we focus on \emph{communication/bandwidth cost}, and assume that synchronization preserves the synchronous ordering of gradient iterations, although our techniques are also compatible with other scheduling strategies, e.g.~\cite{jayarajan2019priority, peng2019generic, jiang2020unified,yi2020fastt}.

\paragraph{Batch Scaling.} 
An orthogonal scaling approach is increasing the batch size at each node. 
This requires careful hyper-parameter tuning for accuracy preservation, e.g.~\cite{goyal2017accurate, you2019large}, although recipes for large batch scaling are known for many popular models. 
We consider scalability in both 
1) \emph{the large-batch setting}, where we adopt the best-known hyperparameter recipes to preserve accuracy, and 2) \emph{the small-batch setting}, corresponding to datasets or models for which large-batch scaling parameters are unavailable or unknown.

% \paragraph{Communication scheduling}
% Another orthogonal idea to tackle communication costs is. 
\subsection{Communication Compression Methods}
\label{sec:methods}

The basic idea behind communication-compression methods is to reduce the bandwidth overhead of the gradient exchange at each step by performing lossy compression. 
Our presentation assumes that a generic mechanism allowing for all-to-all communication among the nodes is available. (We discuss our implementation choices in Sections~\ref{sec:challenges} and~\ref{sec:design}.)
Roughly, existing schemes can be classified as follows. 

\paragraph{Gradient Quantization.} 
This approach works by reducing the bit-width of the transmitted updates~\cite{seide20141}. 
% proposed the first instance of such a scheme, which worked by normalizing gradients before transmission and then rounding the values directly to binary values. These values are then transmitted in compressed form, and decoded by the receiver. Importantly, the compression error  is stored at each node, and added to the gradient of the next optimization step, before compression. 
One of the first compression approaches~\cite{alistarh2017qsgd} observed  that \emph{stochastic} quantization of the gradient values is sufficient to guarantee convergence. 
Their method, called QSGD, is a codebook compression method which quantizes each component of the gradient via randomized rounding to a uniformly distributed  grid. Formally, for any non-zero vector $\vec{v}$, given a codebook size $s$ and $\vec{v} \in \mathbb{R}^d, Q_s(v_i) = \|{\vec{v}}\|_2 \cdot sign(v_i) \cdot q(v_i,s)$. The stochastic quantization function $q(v_i,s)$ essentially maps the component's value $v_i$ to an integer quantization level, as follows. Let $0 \leq \ell \leq s - 1$ be an integer such that $|v_i|/\| {\vec{v}} \| \in [\ell/s, (\ell + 1)/s]$. That is, $\ell$ is the lower endpoint of the quantization interval corresponding to the normalized value of $v_i$. Then,
$$
q(v_i, s) = 
\begin{cases}
\ell/s, \text{with probability } 1 - p(|v_i|/\|{\vec{v}}\|, s), \\
(\ell + 1) / s, \text{otherwise}
\end{cases}
$$
where  $p(a, s) = as-\ell$ for any $a \in [0, 1]$.
The trade-off is between the higher compression due to using a lower codebook size $s$, and the increased variance of the gradient estimator, which in turn affects convergence speed. This idea inspired a range of related work~\cite{lim20183lc, ramezani2021nuqsgd, faghri2020adaptive} reducing the variance of the compression by improved quantizers. We discuss these schemes further in Section~\ref{sec:design}.

% . Specifically, 3LC~\cite{lim20183lc} improved practical compression by combining 3-value quantization with sparsity, quartic encoding, and zero-run encoding. 
% Non-uniform QSGD~\cite{ramezani2021nuqsgd} considered adjusting quantization levels to reduce variance, whereas follow-up work considered learning the locations of the quantization levels~\cite{faghri2020adaptive}. 

\paragraph{Gradient Sparsification.}
 These methods, e.g.~\cite{strom2015scalable, dryden2016communication, lin2017deep, karimireddy2019error}, capitalize on the intuition that many gradient values may be skipped from transmission. 
The standard approach to sparsification is \emph{magnitude thesholding}, effectively selecting the top $K$ gradient components for transmission, where $K$ is a hyper-parameter. 
Then, error correction is applied to feed the thresholded gradient components back into the next round's gradient. 
Variants of this procedure can achieve more than $100\times$ gradient compression while still recovering accuracy~\cite{lin2017deep}. However, this comes at the price of model-specific hyper-parameter tuning, which may be unreasonable in a deployment setting. 

Renggli et al.~\cite{renggli2019sparcml} proposed efficient sparse collectives,  and observed that sparsification methods can be promising in cases where there is high natural redundancy--such as fully-connected or embedding layers--but may be a poor choice for general compression due to the need for  hyper-parametrization. 
Our investigation confirmed their finding. 

% as well as sparsity ratios for which recovery occurs under standard hyperparameters. 
% Generally, they found that recovery is possible for a range of models when the gradient density after thresholding is between 1\% and 10\%.  However, they also noticed that gradient density naturally \emph{increases} during the reduction sum, as the non-zero indices do not necessarily overlap, and that the sparse representation induces additional overheads. 
% Generally, their results suggest that sparsification methods can be promising in very specific cases where there is high natural redundancy--such as fully-connected or embedding layers--but may be a poor choice for the general case. 
% Similar results were recently obtained by~\cite{ramezani2021nuqsgd}.
% Our investigation confirms this finding. 
%Generally, they found that 

\paragraph{Gradient Decomposition.}
This approach treats the gradients as multidimensional tensors, and decomposes the gradient matrix $G \in \mathbb{R}^{m \times n}$ into 2 rank-$r$ matrices $P \in \mathbb{R}^{m \times r}$ and $Q \in \mathbb{R}^{r \times n}$, with $r$ much smaller than $m$ and $n$.
 ATOMO~\cite{wang2018atomo} uses singular value decomposition (SVD) to find the matrices $P$ and $Q$. However, in the case of large models, the SVD of gradient matrices becomes too compute-intensive to be used during training. PowerSGD~\cite{vogels2019powersgd} uses a generalized power iteration algorithm to calculate the matrices $P$ and $Q$, and is the fastest currently-known factorization method. To recover accuracy, it applies a combination of error correction techniques.
Their results show that these methods can be highly useful in the case of CNNs, yielding high compression ratios (up to $100\times$). However, in our experience, recovering accuracy in e.g. Transformers training requires careful tuning, and higher rank values, resulting in lower performance.

% The advantages of these methods persuaded maintainers of PyTorch to integrate PowerSGD method in distributed training library package.
\paragraph{Adapting Compression during Training.} 
The idea of adapting the degree of compression during different stages of DNN training has been considered by  \cite{Guo2020,Chen2018,Chen2020,Abdelmoniem2021}. 
However, we emphasize the fact that all these references in practice \emph{globally adapt} the amount of gradient compression for the entire model to preserve end accuracy, whereas we investigate mechanisms which adapt compression at the per-layer level. Moreover, to achieve high compression, some existing methods require hyperparameter tuning \cite{Chen2020}. A work\cite{Agarwal2021} that supports per-layer compression parameters has a very limited choice of compression parameters (namely picks out of two parameters), requires additional hyperparameter tuning and focuses on specific architectures. By contrast, we adapt compression parameters automatically both across layers, and across training iterations.

\begin{table*}[t]
\centering
\caption{Compression approaches. Stateful here means that approach requires maintaining of a state of error compensating techniques. }
\label{table:compression_approaches}

{\footnotesize
\begin{tabular}{|P{1.8cm}|P{2cm}|P{1.8cm}|P{4cm}|P{1.6cm}|}
\hline
\centering
& Compression rate with recovery & Tunable Parameters & Properties & Computational Overhead \\
\hline
Quantization & $\sim$ 8x & Bits, bucket size & Non-associative, stateless & $\leq$ 3\% \\
\hline
Sparsification (TopK) & $\sim$ 100x & Sparsity, momentum & Non-associative, stateful, not overlapping with compute & 10\%\\
\hline
Decomposition (PowerSGD) & $\sim$ 100x & Rank, warm-up & Associative, stateful, incompatible with mixed precision & 20\% \\
\hline
\end{tabular}
}
\end{table*}

\paragraph{Efficient Software Support.} 
There has already been significant work on providing system support for  compression. Two main challenges are: 1) the introduction of additional hyper-parameters in the training process, and 2) the fact that, since most compression methods are \emph{not associative}, they are not directly supported by standard collective implementations and require algorithm-specific re-implementations.
Grubic et al.~\cite{grubic2018synchronous} showed that  CNNs can withstand 8-bit gradient compression, and provided a simple MPI-based implementation of quantization, while Dutta et al.~\cite{dutta2020discrepancy} examined the implementation gap, showing that frameworks should support both global and per-layer compression. Renggli et al.~\cite{renggli2019sparcml} and Fei et al.~\cite{fei2020efficient} provided efficient support for sparse reductions, while the GRACE framework~\cite{grace}, Bagua~\cite{gan2021bagua} and HiPress~\cite{bai2021hipress} frameworks provided efficient implementations of communication-compression methods. We compare against these frameworks in Section~\ref{sec:experiments}. 

We differ from this prior work in two major directions. At the application level, we focus on \emph{seamless, parameter-free integration} with existing data-parallel training pipelines: thus, we investigate compression techniques which allow accuracy recovery \emph{without additional hyper-parameter tuning}. This is not the case with prior frameworks, which leave the choice of compression parameters to the user. 
Second, at the system level, we seek to maximize speedup by rewriting components of the communication stack to support compression, provide an adaptive layer-wise compression solution which maximizes speedup. 

Recent work by~\cite{agarwal2021utility} investigated the practical potential of gradient compression methods in cloud-grade settings.  They provide analytical and empirical evidence suggesting that gradient compression methods can only provide marginal speedups in distributed data-parallel training of DNNs in such bandwidth-overprovisioned settings. 

However, the generality of their results is restricted by the following factors: 
1) they only consider a limited subset of compression methods and possible implementations: for instance, their compressed implementations strictly follow the NCCL API, which, as we illustrate via our QNCCL implementaiton, means that the compression methods were used inefficiently and with accuracy loss; 
2) they focus on cloud-grade bandwidth-overprovisioned systems, and therefore their findings do not apply to the popular setting of commodity servers. 
These two factors, as well as additional implementation differences, explain the difference between their conclusions and the ones from this work.

% our focus is on determining whether communication-compressed methods are a viable alternative to hardware-level approaches for bandwidth-bound ML workloads, . To our knowledge, this question has not been investigated in prior work, which focused either on building support for specific method families (e.g.~\cite{grubic2018synchronous, renggli2019sparcml, fei2020efficient}) or on using compression to complement bandwidth over-provisioning for cloud-grade hardware~\cite{grace}. 

\section{Goals and Challenges}
%Here we show some numbers 
\label{sec:challenges}

The results in Section~\ref{sec:motivation} suggest that bandwidth can be a key bottleneck when attempting to scale DNN training on commodity GPUs, while the discussion in Section~\ref{sec:methods} outlines non-trivial trade-offs when implementing these techniques for general models. 
We therefore outline our key goals: 
\begin{enumerate}
    \item \textbf{Accuracy Recovery:} Similar to MLPerf~\cite{mattson2020mlperf}, we set our accuracy loss threshold at $<1\%$ relative to the main metric of the full-precision baseline (e.g. Top-1 classification accuracy), although in most of the tasks we present the accuracy loss is practically negligible.  

    \item \textbf{Hyperparameter-Freedom:} Second, we wish to enable scalable data-parallel DNN training in the absence of any model or task information, recovering accuracy under \emph{standard (uncompressed) hyper-parameters}. 
    
    \item \textbf{Eliminating Bandwidth Bottlenecks:} Third, we aim to mitigate or even completely eliminate bandwidth constraints. 
Since not all target models are equally communication-bottlenecked, this allows us some flexibility with respect to how much compression to apply depending on the model and application. 
\item \textbf{Simple Interface:} Finally, the integration with the underlying training framework should be seamless. 

\end{enumerate}

\paragraph{State of the art.} We executed implementations of the compression methods described in Section~\ref{sec:methods} on a range of modern tasks and models. Our findings are summarized in Table~\ref{table:compression_approaches}, and discussed in detail below.  

\emph{We found that no existing approach fully satisfies all the above requirements.} For instance, \emph{quantization-based methods} are known recover accuracy on CNNs when using 8-bit compression~\cite{grubic2018synchronous}, meeting Goals 1 and 2. However, this amount of compression is not sufficient to remove the bandwidth bottlenecks for modern Transformer-class models (Goal 3); moreover, the parameters of~\cite{grubic2018synchronous} do not allow full accuracy recovery on Transformers.

Second, examining \emph{gradient sparsification} methods, we notice that they can ensure high compression (Goal 3); however, they require complex hyperparameter tuning for accuracy recovery in the high-compression regime~\cite{lin2017deep}, breaking either Goal 1 or Goal 2. Conversely, as also noted by~\cite{renggli2019sparcml}, these methods can recover accuracy under medium density (e.g. 20\%), but in that case their performance is similar to quantization approaches. This family of methods has the additional cost of having to maintain state (the error buffer) and being less amenable to computation-communication overlap, since the selection operation is applied over the entire gradient.  

Finally, \emph{decomposition} methods have been shown to yield compression ratios of up to $100\times$ in the case of CNNs, attaining Goal 3. Moreover,  with careful tuning of hyper-parameters, PowerSGD is able to recover accuracy for CNNs under generic rank-decomposition values. In addition, this method is \emph{associative}, lending itself to seamless implementation via MPI or NCCL (Goal 4). 
Unfortunately, however, we found that this method can require high rank values for stable training, especially on  Transformers, where there is almost no speedup, and that it is not compatible with reduced-precision (FP16) training, which is used by  virtually all frameworks.

\begin{figure}
     \centering
     \includegraphics[width=0.3\textwidth]{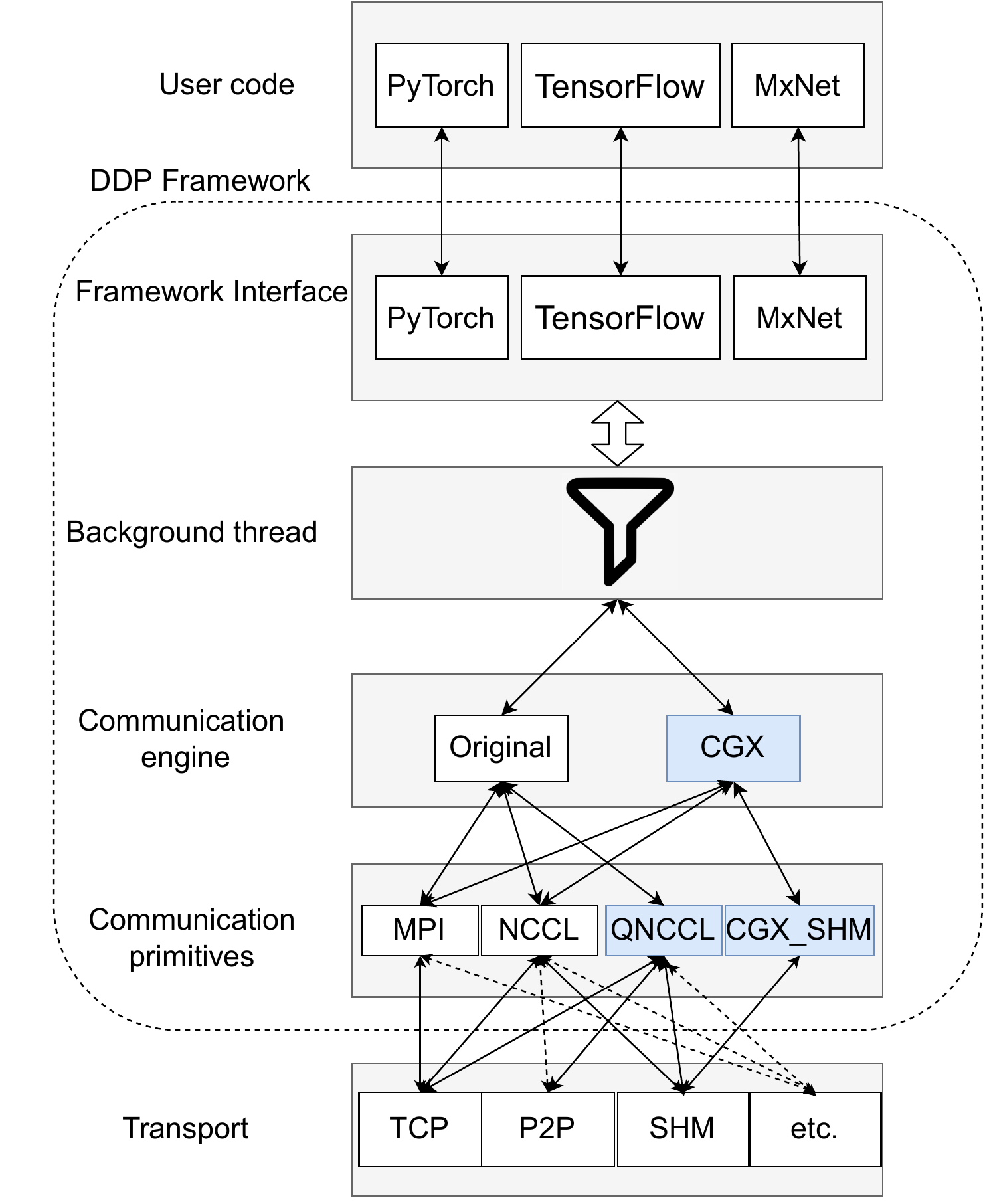}
     \caption{Abstract architecture of a Distributed Data Parallel (DDP) framework. CGX components are in blue, and arrows stand for procedure calls. Dashed arrows represent hardware interactions, e.g. P2P transport is supported via GPU NVLinks.}
     \label{fig:design}
 \end{figure}

\section{CGX System Design}
%Here we describe the design of the system and explain how the design choices have been made
\label{sec:design}
\paragraph{ML Frameworks under the Hood.}
A typical DNN training framework has three parts, as described in Figure~\ref{fig:design}:
% reference to a figure with Design
\begin{enumerate}

\item \textbf{Framework interface} (in Python) with high-level API is called by User code. It may also include a frontend that unifies the input from the learning framework. 

\item \textbf{Background thread} collects inputs, groups them into blocks based on query type and input properties, schedules the reduction for each block.

\item \textbf{Communication engine} performing the query (Allreduce, Broadcast, Allgather). At this stage, the framework typically calls an existing communication library, such as NCCL (QNCCL), Gloo, or an MPI implementation.

\end{enumerate}

A key issue when implementing most compression methods such as quantization or sparsification is that their operations are \emph{non-associative}, and so the aggregation function (sum) must be performed {at the lowest level} in the above diagram. This means that we cannot integrate the compression into higher levels without a bespoke implementation, which in turn may lead to performance and implementation costs.

\subsection{The CGX Communication Engine} 

To efficiently support compression, we implemented our own  communication engine (blue component on Figure~\ref{fig:design}), with  primitives which support non-associative compression operators. Broadly, there are two approaches to do this. The first is a \emph{native} one, by which one can implement compression-aware Allreduce  using communication libraries. Alternatively, one can modify or extend existing communication libraries, such as NCCL, to support compression operators. 

The native approach requires deeper integration, but has the advantage that compression is performed ``closer" to training, which means that the compression engine has information about the model layers, and their gradients and thus has a richer, more flexible API. The disadvantage is that it has to explicitly interface with the training framework, and users may have to adjust their training pipeline. 

The second \emph{low-level} approach is to directly perform compression and de-compression at the primitive/transport level, independently of the user's code and training pipeline. In this case, the framework can only operate with the raw data buffers provided by the upper layers. This loses information about the data it operates with, e.g., layer names, which could be useful for compression operators, but is easier to interface with, and may have lower overheads. 

\subsubsection{Framework Integration}
To investigate this non-trivial dichotomy, we implemented \emph{both} variants. 
Specifically, our main framework, called CGX, integrates natively with the user's code, and can interface both via Horovod~\cite{sergeev2018horovod}, a popular distribution wrapper that works with all major ML frameworks, but also separately via framework-specific extensions, such as PyTorch Distributed Data Parallel (DDP). Separately, as an instance of the ``low-level'' approach, we re-implemented the NCCL communication library to support quantized reduction operations. We call this separate implementation QNCCL, and contrast it to our main approach.

\paragraph{The Native CGX Framework.}
\label{sec:layer_filters}
The main version of CGX uses the  Horovod wrapper~\cite{sergeev2018horovod} to interface with popular ML frameworks. Specifically, we implemented a communication engine with Allreduce methods supporting compression operators. 
Next, we added layer filters that split model gradients into logical subsets, which the framework may handle differently: some accuracy-critical subsets are communicated in full precision, while other subsets are compressed and reduced in lower-precision. 
Empirically, it is known that layers like batch/layer normalization and bias layers are sensitive to gradient compression, while being small. Therefore, we communicate them uncompressed. As a bonus, this avoids calling compression operators for multiple small inputs.
At the filtering level, the framework also performs packing or splitting of levels into the units of communications, so called fused buffers. Typical size is around 64MB. The communication engine then performs reduction with these units not the layers. But it keeps the information of offsets of the layers within the fused buffers because this information will be used for layer-wise compression.

Further, CGX performs compression \emph{per-layer}, and not as a blob of concatenated tensors. This provides the flexibility of exploring heterogeneous compression parameters and avoids mixing gradient values from different layers, which may have different value distributions, leading to large quantization error. We found that such filters can be applied ``at line rate'' without loss of performance, as most of the computation can be overlapped with the transmission of other layers. 
CGX's API allows users to choose the compression parameters for specific layers or filter out the group of layers. 

\paragraph{Torch DDP Integration.} Our compression/communication engine is portable: to illustrate this, we also integrate it separately with the Torch DDP pipeline~\cite{paszke2019pytorch}. 
% The code is available at \textbf{\url{https://github.com/IST-DASLab/torch_cgx}}. 
In this case, CGX acts as a Torch extension that implements an additional Torch DDP backend, as a supplement to the built-in NCCL, MPI and Gloo backends. Thus, users only need to import the extension and change the backend at initialization. 

We integrated our functionality into the communication engine of the Data Parallel framework. At this level, we no longer have access to the buffer structure, therefore we can not explicitly filter layers. Nevertheless, the user can provide  the layout of the model layers (e.g. gradient sizes and shapes). Using this information, we can obtain the offsets of the layers in each buffer provided by \texttt{torch.distributed}.
%An example of using the Torch extension is presented in the Appendix.

\subsubsection{Choosing a Reduction Scheme}
% \label{sec:implementation-details}
% \subsubsection{Communication}
% \label{paragraph:backends}

% The key question at the immediate lower level of the communication stack is the Allreduce implementation.
% For this, we considered implementations supporting GPU-based inputs and providing peer-to-peer primitives. Here, the main existing options are GPU-aware MPI implementations, or NCCL. 
% Instead of relying on existing work, we decided to implement 

% As an alternative, we implemented a set of new peer-to-peer communication primitives that are based on data transfers through UNIX shared memory %~\cite{UnixSM}.
% We call this backend SHM.

% \paragraph{Reduction Schemes.} 
The ``hottest'' operation in distributed data-parallel training is Allreduce, corresponding to the logical gradient averaging. To support non-associative compression operators, we need to choose the reduction algorithm together with the compression operator, to maximize performance and minimize the  compression error due to iterative compression-decompression. We considered the following reduction schemes. 

 \noindent \textbf{Scatter-Reduce-Allgather (SRA)}  works in two rounds: a  process first divides its vector of dimension $d$ into $N$ subarray ``chunks;'' each node receives its chunk of the initial vector from all other nodes and aggregates it (Scatter-Reduce). Second, it broadcasts the aggregated chunk (Allgather). The bandwidth cost is $O(d(N-1))$, the latency term is $2\alpha$, corresponding to the two rounds.  \textbf{Ring-Allreduce} is the bandwidth-optimal algorithm, implemented in most libraries (e.g. NCCL, Gloo). Similar to SRA, it divides the initial vector into chunks, and communication is done in a ring-shaped topology. In the first phase, each node sends a chunk to its ``right'' neighbor and receives a chunk from its left neighbor. It then sums the received chunk with its local result and sends the result forward, repeating $N - 1$ times. 
    In the second phase, nodes broadcast (Allgather) the resulting chunks on the ring. 
    The bandwidth cost is $O(d(N - 1) / N)$, with latency $2 \alpha (N - 1)$, assuming communication can not be itself parallelized.
    \textbf{Tree-Allreduce} can be seen as a hierarchical parameter-server. Communication is done in $2 \log N$ rounds and two phases. The nodes build a tree-like topology, and send their vectors up to the root, summing them along the path, and then propagate back the result. Communication complexity is $O(2d \times \log(N))$, while latency is  2$\alpha \log N$.
    
\paragraph{Discussion.} 
We examined the practicality of these reductions, and found \textbf{Scatter-Reduce-Allgather (SRA)}  to show the best performance. It also has the key algorithmic advantage of \emph{lower compression error}, due to fewer compression/decompression steps. Thus, we mainly employed this algorithm inside CGX. 
Table~\ref{table:reductions} illustrates CGX throughput under different reduction schemes, for different tasks, on an 8-GPU server. (See Section~\ref{sec:experiments} for the full setup.) 

\begin{table}[h]
\centering
\caption{Throughput of different reduction schemes (items per second).}
\label{table:reductions}
{\footnotesize
\begin{tabular}{|c|c|c|c|}
\hline
 &  ResNet-50  & Transformer-XL  & ViT  \\
\hline
SRA & \textbf{2900} & \textbf{260k} & \textbf{1918} \\
\hline
Ring & 2830 & 236k & 1883 \\
\hline
Tree & 2770 & 202k & 1756 \\
\hline
\end{tabular}
}
\end{table}

\subsubsection{Default Compression Approach} 
Our framework implements several compression approaches; yet, based on the discussion in Section~\ref{sec:methods}, we use \emph{gradient quantization} as our main method. 
The rationale behind our choice is the following. 
First, as suggested by Figure~\ref{fig:fake_compression}, quantization compression by 8-10x should provide sufficient bandwidth reduction to overcome most of the communication bottleneck. 
Moreover, it can do so \emph{in a generic, parameter-free way}: an independent contribution of our work is that we identify  
\emph{general parameter values providing 8-10x compression without accuracy loss on all the model classes and tasks we tried}. We investigate additional performance improvements customized per-layer compression, which can provide an additional performance boost.

\begin{figure}
     \centering
     \includegraphics[width=0.5\textwidth]{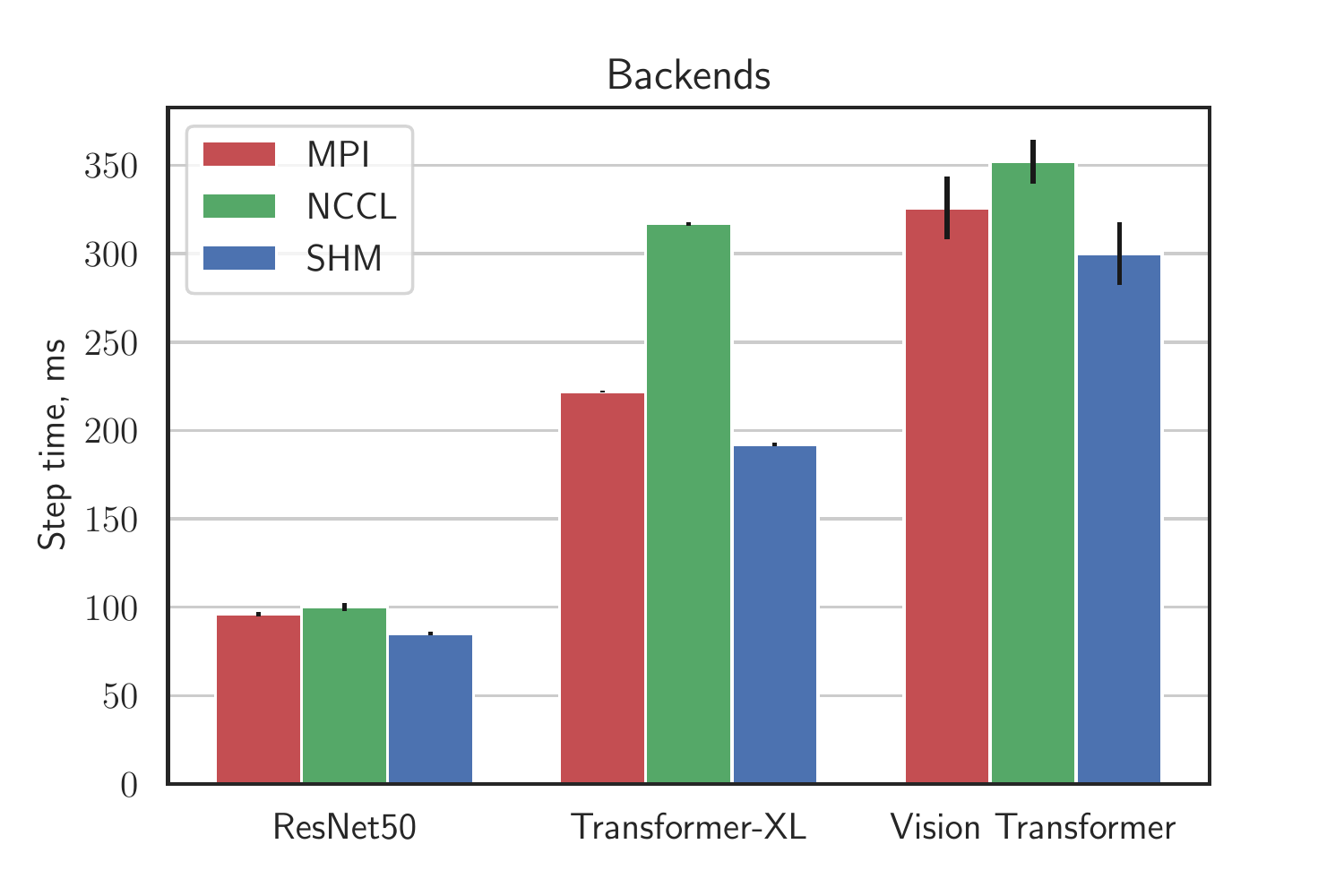}
     \caption{Training step times for different communication backends in CGX Communication engine on a single node, 8 RTX3090 GPUs. Lower is better.}
     \label{fig:backends}
 \end{figure}

\begin{figure}
    \centering
    \setkeys{Gin}{width=0.49\linewidth}
    \subfloat[Communication latency.\label{fig:backends_lat}]{\includegraphics{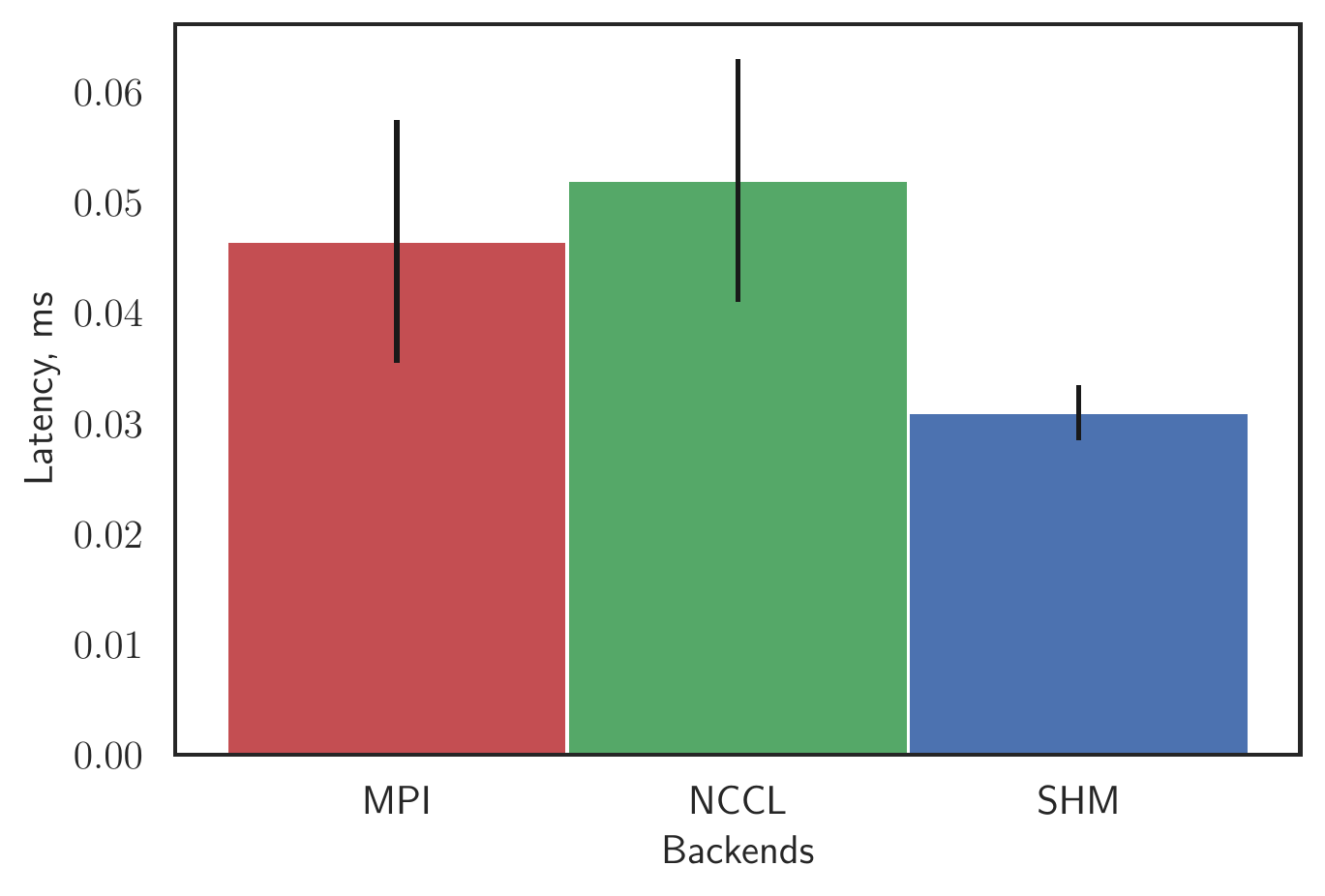}}\hfil
    \subfloat[Communication time.\label{fig:backends_hop}]{\includegraphics{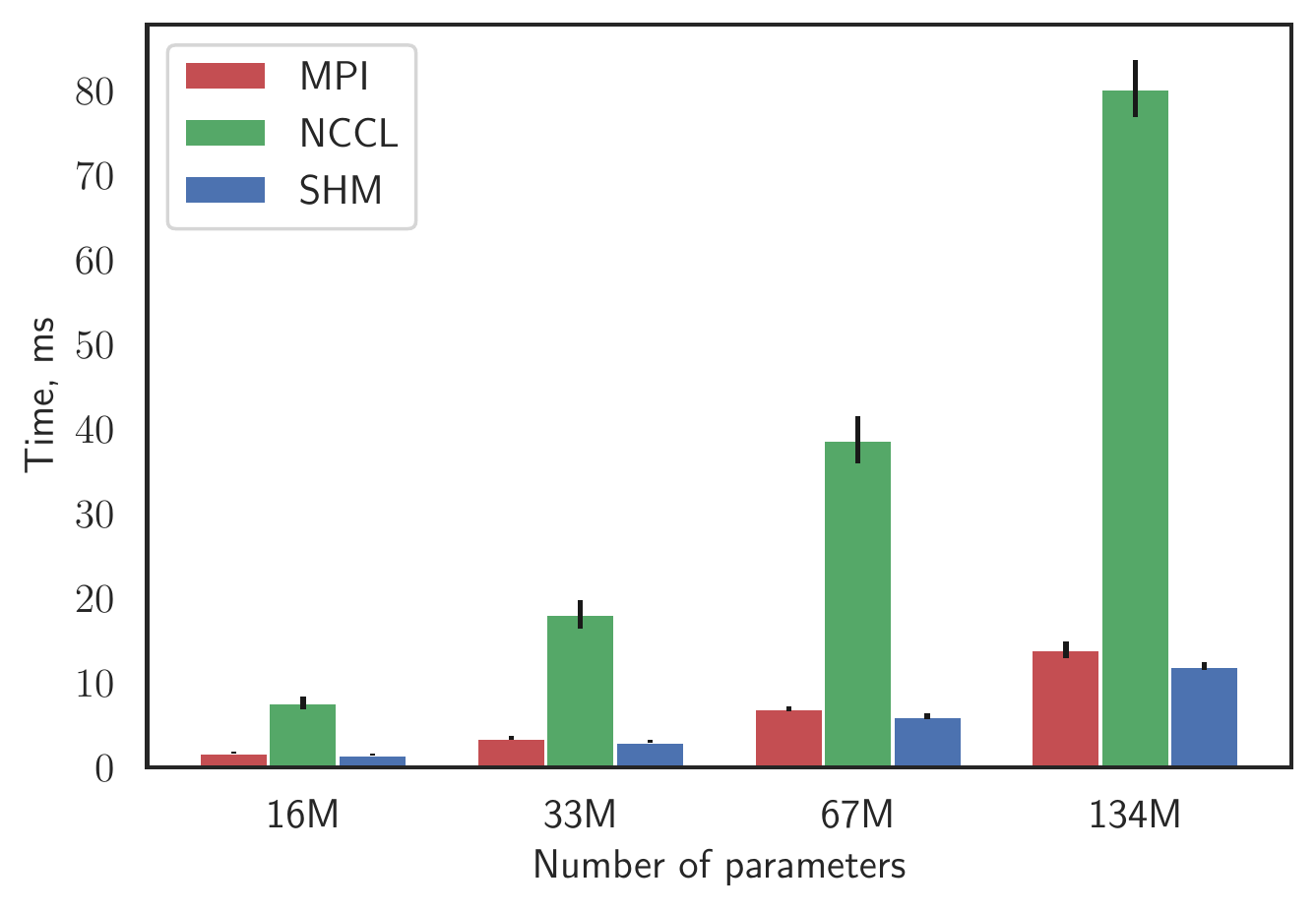}}
    \caption{Comparison of point-to-point communication using different backends.}
    \label{fig:backends_p2p}
\end{figure}

\subsection{Communication Backend}

The key question at the lower level of the stack is how to implement the point-to-point communication primitives. Here, existing options are GPU-aware MPI implementations, NCCL, or Facebook Gloo. (For instance, GRACE~\cite{grace} supports all three options.) 
To maximize performance, instead of relying on existing implementation, we developed a set of new point-to-point communication primitives, that are based on data transfers through UNIX shared memory. We call this communication backend SHM.

% As an alternative, we implemented a set of new peer-to-peer communication primitives that are based on data transfers through UNIX shared memory %~\cite{UnixSM}.
% We call this backend SHM.

SHM works by registering a UNIX shared memory buffer for each pair of GPUs within a node and mapping it to GPU memory. On send, we move the input buffer to the shared segment and synchronize with the recipient using CUDA IPC primitives. At SHM communicator initialization, we allocate an 2 auxiliary buffers for each one directional point-to-point communication. The size of the buffer is the size of the fused buffer, i.e. 64 MB.  It means that for 8 GPU communication (e.g. for SRA all All-to-All communication) we allocate $128 * 7 = 896MB$ on each GPU.
SHM is only supported for a single server, while CGX can use both MPI- and NCCL-based backends in multi-server setups. Moreover, we support heterogeneous communication where the intra node communication uses SHM, MPI, or NCCL as the backend, or performs NCCL-allreduce without compression, while the inter-node communication uses MPI or NCCL.
The difference in performance is illustrated in Figures~\ref{fig:backends} and ~\ref{fig:backends_p2p}. The speedup is justified by the lower synchronization between compression and communication, and the  memory transfers via the GPU Communication Engine. In  Figure~\ref{fig:backends_lat} we show timings for point-to-point communication of small buffers, whereas the Figure~\ref{fig:backends_hop} shows the communication time dependency on large buffers sizes. The figures demonstrate that SHM significantly outperforms other backends.
Thus, unless otherwise stated, we use SHM for intra-node communication in all our experiments.

\subsection{Implementation Details}

\paragraph{Efficient Quantization.}
%TODO buckets, vectorization, compression overhead
The quantization algorithm sketched in Section~\ref{sec:methods} has the following downside: when applied to the entire gradient vector it leads to convergence degradation, due to scaling issues. A common way to address this is to split the vector into subarrays, called buckets, and apply compression independently to each bucket~\cite{alistarh2017qsgd}. This approach increases the compressed size of the vector because we have to keep scaling meta-information for each bucket and slows downs the compression, but helps to recover  full accuracy. The bucket size has an impact on both performance and accuracy recovery: larger buckets lead to faster and higher compression, but higher per-element error. Therefore, one has to pick the bucket size appropriate for the chosen bits-width empirically. We found out that 4 bits and 128 bucket size always recovers full accuracy, has reasonable speedup, and can be efficiently implemented, so we use this as a compression baseline in all our experiments.

% Another important factor is the processing time of compression. In all data-parallel implementations, communication can be overlapped with computation, i.e. the expensive forward-backward operations. This also implies that compression may compete with computation for GPU resources. Therefore, we need to estimate the overhead we have due to extra computation. 
To achieve low compression overheads, we applied the following optimizations: we use an efficient parallel bucket norm computation algorithm, and, for   elementwise compression/decompression, we perform cache-friendly vectorized memory load/stores. 
Quantization overhead amounts to 1-3\% of computational cost in our benchmarks.

% To estimate the compression overheads, we turned off the compression function and communicated a data buffer of the same size as we would receive after the compression. We then compared performance on Transformer-XL and Vision Transformer models, against the experiment with quantization enabled (quantization bits = 4, bucket size = 128). 

\paragraph{Improved Scheduling.}
CGX also aims to improve the latency term. For this, we perform  fine-grained scheduling of gradient synchronization, which is known to lead to improved performance for Parameter Servers~\cite{jiang2020unified}. The scheduling of the communication is task-based, where each task are layer gradients that we want to synchronise every iteration. In the background thread we collect the tasks until the total size of collected gradients reaches user-defined size $B$ or user-defined cycle time $C$ expires. Then the concatenated group of gradients is synchronised. The constants $B$ and $C$ are autotuned; we leveraged parts of this implementation from Horovod and \texttt{torch.distributed}. As part of scheduling optimization, CGX supports user-defined filtering of layers and cross-barrier training. Filtering of small layer modules such as biases or batch norm not only improves convergence, but positively affects performance. Such filtering removes the need of extra compression kernels calls without notable increase of communication costs. Cross-barrier optimization does not provide significant performance in a single node setup, confirming the  observations in~\cite{jiang2020unified}.

%(see Figure~\ref{fig:txl-metric}). 

\subsection{The QNCCL Library}

The role of the QNNCL implementation is to contrast our design choices relative to a direct re-implementation of communication compression in the popular NCCL library. To build this low-level variant, we started from vanilla NCCL and replaced Allreduce with implementations that compress every piece of data before its transfer. Basically, in DDP stack(Figure. ~\ref{fig:design}) we replace NCCL with our version that supports compression.
%The code is available at \textbf{\url{https://github.com/IST-DASLab/QNCCL}}.
We leverage the NCCL communication optimizations, to avoid costs for additional GPU calls. However, in this case, we lack information about the internal structure of the buffer, and have to apply compression parameters uniformly over the entire model. 
In this case, we also have limitations in terms of the GPU resources imposed by NCCL itself, which lead to additional compression overheads. 
We examine the performance trade-offs of this approach in the experimental section. 

\section{ Layer-wise Adaptive Quantization}

One key optimization supported by CGX is \emph{varying compression parameters at the per-layer level}. This is especially well-suited to models such as Transformers which have heterogeneous layer sizes, e.g. due to large embeddings. Synchronization of such layers can be quite expensive, and, since they come early in the model, cannot be overlapped with computation. Yet, these massive layers can support highly-compressed communication. 
Thus, we investigate  \textit{automatic} mechanisms to pick  per-layer compression levels.

%, or we can even apply gradient sparsification. 

% As an optimization target, we picked the Transformer-XL model, which is the most heterogeneous model, in terms of layer sizes. In particular, the model has a large first ``embedding'' layer which contains more than 60\% of all model parameters. The synchronization of this layer is the most expensive communication step, and cannot be overlapped with computation as it is the last layer for the backward operation. 
% We have varied the bit-width used to encode gradients for this layer between 1--3 bits, as well as the bucket size between 32--128, leaving all other layers at the default 4 bits and 128 bucket size.

% We found that, if we apply 2 bits compression to the layer, we can recover accuracy with only small bucket sizes ($\leq 32$) which slows down training due to the increase in compression cost. With 3 bits and 128 bucket size, we recover full accuracy and obtain an additional $\approx$ 5\% speedup. When applying the same parameters to all other layers we observed significant accuracy degradation.

% \todo[inline]{The description of the automatic algorithm should be moved to the earlier section.
% I wanted it to logically follow from the problems we encounter. I don't see it in the Design or Implementation sections.
% }

We focus on the trade-off between two parameters for each layer: the \emph{magnitude of the compression error} and \emph{compressed size of the layer}. Our adaptive algorithm tries to balance these constraints in order to maximize speedup while recovering convergence. We  periodically collect gradient statistics  and then re-assign bit-widths and bucket-size to each layer. 
Specifically, we want to minimize the compressed size of the model gradients, while minimizing the $\ell_2$-norm of the compression error, which is linked to convergence~\cite{karimireddy2019error}. 

\paragraph{Problem Definition.} 
We formalize this problem as identifying per-layer bit-widths $b_1, b_2, \ldots, b_L$ for the $L$ layers minimizing the \emph{bandwidth objective} $\sum_{\ell = 1}^L b_\ell \cdot size(L_\ell)$ across all the ${b_i}$s, \emph{subject to} the fact that compression error cannot not exceed a maximum threshold $\alpha \cdot E_4$. Here, $\alpha > 0$ is a fixed parameter, and $E_4$ is the error when we compress all layers to $4$ bits, for which we know that full recovery occurs.

We emphasize that this formulation is different from the (global) adaptive compression problems considered by prior work~\cite{Agarwal2021,Guo2020,Chen2018,Chen2020,Abdelmoniem2021}, as they usually consider the problem of adapting the global degree of compression to the various stages of the training process, as opposed to optimization of the fine-grained layer-wise bit-width adaptation we consider. 

This constrained optimization problem can be approached via standard solvers, and in fact our first approach has been to use Bayesian optimization. 
However, we found that this requires instance-specific tuning, and adds hyper-parameters. 
We therefore investigate problem-specific heuristics. 

A straightforward such approach is to simply sort layers by the ratio of gradient magnitude over the layer size. 
We then assign the lowest bit-width to the first layers in this order, and the highest to the last layers, interpolating linearly in the middle. Experimentally, this approach recovers accuracy and improves over static assignment, but the performance gains are minor.

% optimization problem where we pick discrete quantization levels per layer, minimizing a composite objective balancing the norm of the quantization error with the total number of bits employed per layer. We have  

% We solved this using Bayesian optimization approach. 

% The use of $\ell_2$ error norm is justified theoretically, as this is linked to convergence~\cite{karimireddy2019error}. 

This observation inspires a \emph{clustering-based} approach, by which we collect layers with similar sensitivity to gradient compression into groups, and assign bit-widths correspondingly. 
We use a 2D-clustering algorithm~\cite{Macqueen67somemethods}, where the dimensions are the size of the layer, and the $\ell_2$-norm of the top values of the accumulated gradient. We perform clustering to obtain ``sensitivity groups,'' each with its own centroid, and then sort the centroids by their gradient norms. Finally, we linearly map bit-widths and bucket sizes to the layers. The exact procedure is described in  Algorithm \ref{algo:kmeans_based}. We investigate its practical performance in Section~\ref{sec:adaptive}.

\section{Experimental Validation}
\label{sec:experiments}

% TODO Ilia: Explain the choice of training experiments. Models, batch sizes and mixed precision.
\subsection{Experimental Setting}

\paragraph{Infrastructure.} Our evaluation uses commodity workstations based on RTX2080 and RTX3090 consumer-grade GPUs, and a cloud-grade EC2 \texttt{p3.16xlarge} machine, with 8 V100 GPUs, equivalent to a DGX-1 server. 
Please see Table~\ref{table:systems_char} for complete system characteristics. 
% The scheme of interconnect between GPUs on the 8x RTX3090 machine is shown in Appendix.
In brief, the 8 GPUs are split into two groups, each assigned to a NUMA node, which are bridged via QPI.
Bandwidth measurements via~\cite{li2018tartan} show that inter-GPU bandwidth varies from 13 to 16 GBps depending on location. 
At the same time, we have 1GBps Allreduce bandwidth for reasonable buffer sizes. Results for RTX2080 are similar, with 1.5GBps Allreduce bandwidth.

The V100/DGX-1 machine forms a so-called \textit{Backbone Ring} inside a \textit{Hypercube Mesh}~\cite{DGX_topo}, in which GPUs are connected via NVLINK. 
The DGX-1 has GPU-to-GPU bandwidth of up to 100 GBps, leading to the same Allreduce bandwidth our workloads. Performance on our setup is identical to a branded DGX-1 measured via NVIDIA's  benchmarks~\cite{DeepLearningExamples}.

\paragraph{Environment and Tasks.}
Most experiments were run using the PyTorch version of the NVIDIA Training Examples benchmark~\cite{DeepLearningExamples}. For state-of-the-art model implementations we used the Pytorch Image Models~\cite{rw2019timm} and the Huggingface Transformers repositories~\cite{HF}. For the experiments on V100 machine we used the official  NGC PyTorch 20.06-py3 Docker image. We used CUDA 11.1.1, NCCL 2.8.4, and cudnn/8.0.5.
We examine three different DNN learning tasks:
1) image classification on ImageNet~\cite{deng2009imagenet} ; 2) language modeling on WikiText-103; 3) question-answering on the SQUAD dataset.

\begin{algorithm}[t]
 \caption{KMEANS-based adaptive compression}
 \label{algo:kmeans_based}
 {\small
 \begin{algorithmic}[1]
 \renewcommand{\algorithmicrequire}{\textbf{Input:}}
 \renewcommand{\algorithmicensure}{\textbf{Output:}}
 \REQUIRE Model Layers $L_i$, accumulated gradients $G_i$, possible bit-widths $B = \{\beta_1, \beta_2, \ldots, \beta_k\}$
 \ENSURE Bit-width assignments $b_\ell \in B$ for each layer $\ell$\\
 \textit{Initialisation} : Compute 2D-representation for each layer $\ell$ by computing points ($size(L_\ell)$, $norm(G_\ell)$).
  \STATE Obtain (centroids, clusters) = kmeans over data into $k$ clusters
  \STATE Sort centroids based on $norm(C_i) - size(C_i)$ and assign them
  \STATE Assign points (layers) corresponding to each centroid to the corresponding bit width $b_\ell$.
 \end{algorithmic}
 }
\end{algorithm}

\begin{table*}[t]
\centering
\caption{Validation results for training with the baseline and CGX optimizations, respectively. ResNet50, VGG and ViT numbers are Top-1\% accuracies, Transformer-XL and GPT-2 show perplexity, while BERT shows F1-score.}
\label{table:val_results}
{\small
\begin{tabular}{|c|c|c|c|c|c|c|}
\hline
\centering
& ResNet50 & VGG16 & ViT-base & Transformer-XL-base & GPT-2 & BERT\\
\hline
Baseline & $75.8 \pm 0.2 $ & $69.1 \pm 0.1$ & 79.2  & $22.81 \pm 0.1$ & $14.1 \pm 0.1 $ & $93.12 \pm 0.05$ \\
\hline
CGX  & $75.9 \pm 0.2$ & $68.9 \pm 0.1$ & 78.6 & $22.9 \pm 0.1 $ & $13.9 \pm 0.1$ & $93.06 \pm 0.05$\\
\hline
\end{tabular}
}
\end{table*}

\paragraph{Baselines.} We use the non-compressed original training recipes as a baseline. We \emph{do not} modify any of the training hyper-parameters. In distributed training, we use either Horovod-NCCL or PyTorch-DDP with NCCL backend. In all our experiments, NCCL showed better performance than OpenMPI or Gloo, so we use it as the default backend. For a fair comparison,  we use the CGX extension depending on the baseline framework: for Horovod-NCCL, we use our Horovod extension, and for  PyTorch-DDP we apply our Torch distributed backend extension. 
We also compare our results against ideal linear scalability on the same machine, calculated by training speed on a single device  multiplied by the number of devices. We use step time and throughput (items/sec) as the performance metrics.
For all performance experiments, we validated that the hyper-parameters used are sufficient to recover training accuracy, across 3 runs with different seeds. All the reported speed numbers are averaged over 300 training iterations after a warm-up of 10 iterations.
Unless specifically stated, we \emph{do not} employ the adaptive compression algorithm.

\begin{figure*}[h]
    \centering
    \setkeys{Gin}{width=0.33\linewidth}    
    \subfloat[ResNet50, ImageNet\label{fig:throughput-rn50}]{\includegraphics{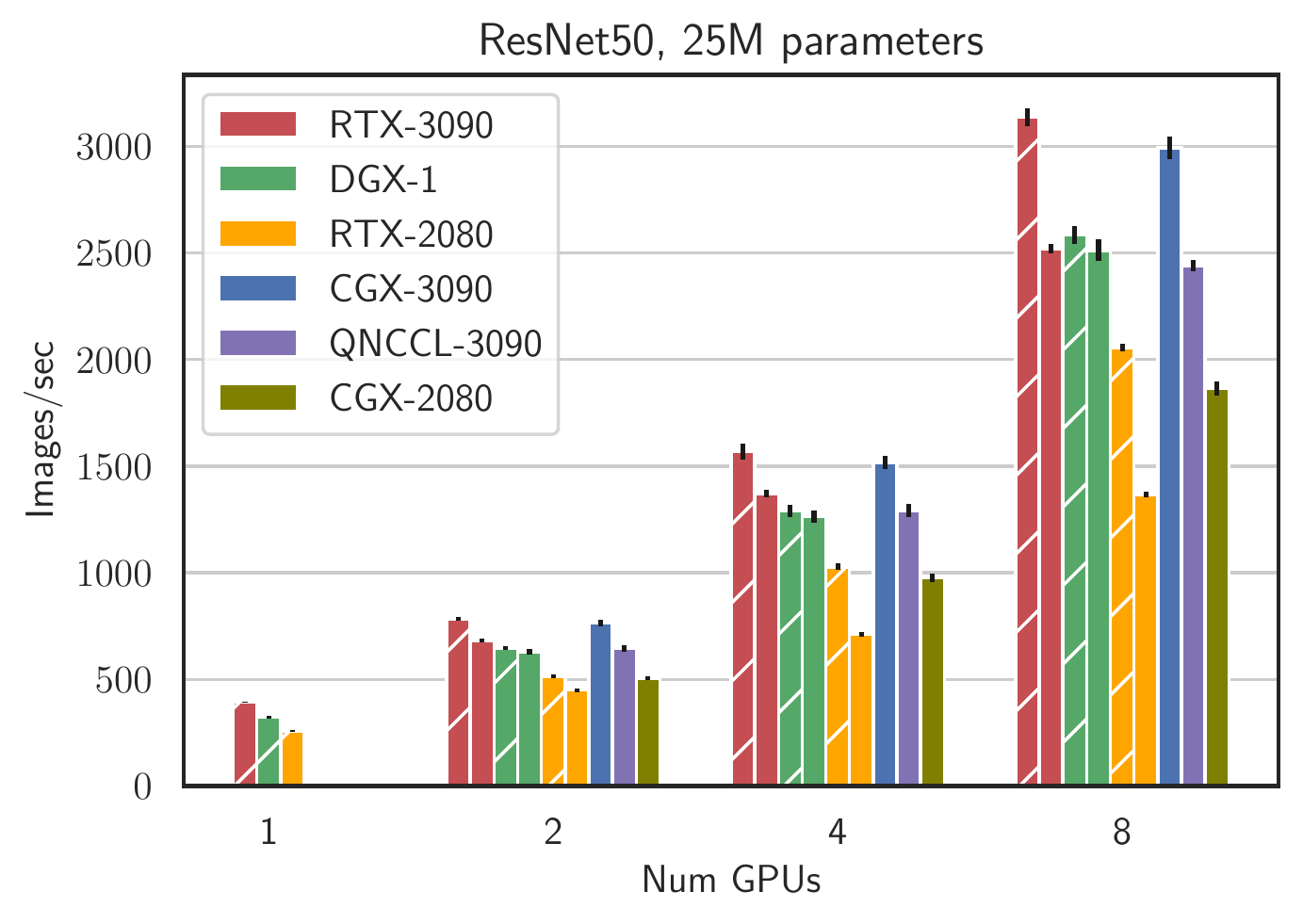}}% \hfil
    \subfloat[Transformer-XL, WikiText103\label{fig:throughput-txl}]{\includegraphics{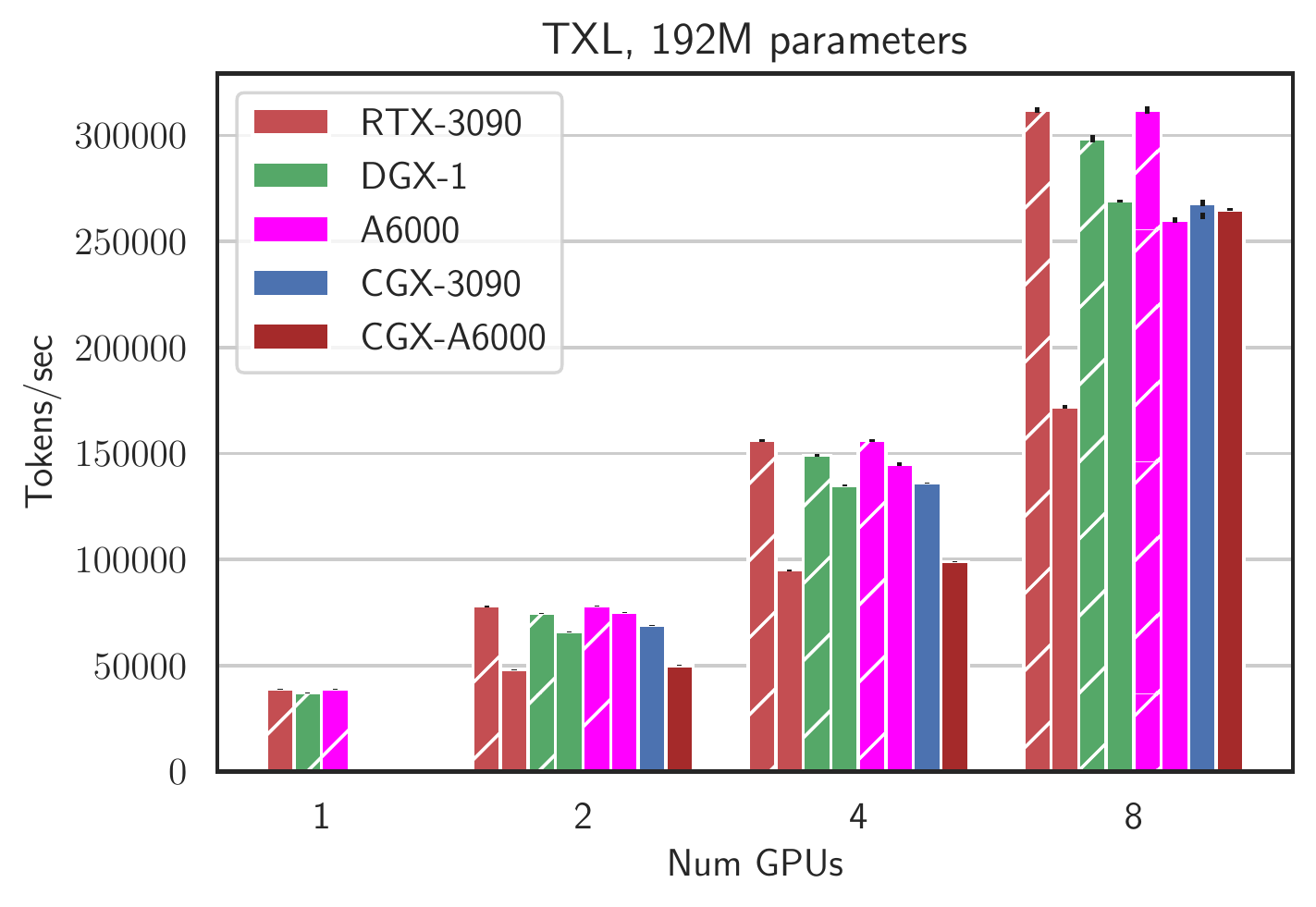}} % \hfil
    % \subfloat[Vision Transformer, ImageNet\label{fig:throughput-vit}]{\includegraphics{images/Vit_comparison.pdf}} % \hfil
    \subfloat[BERT, SQUAD\label{fig:throughput-bert}]{\includegraphics{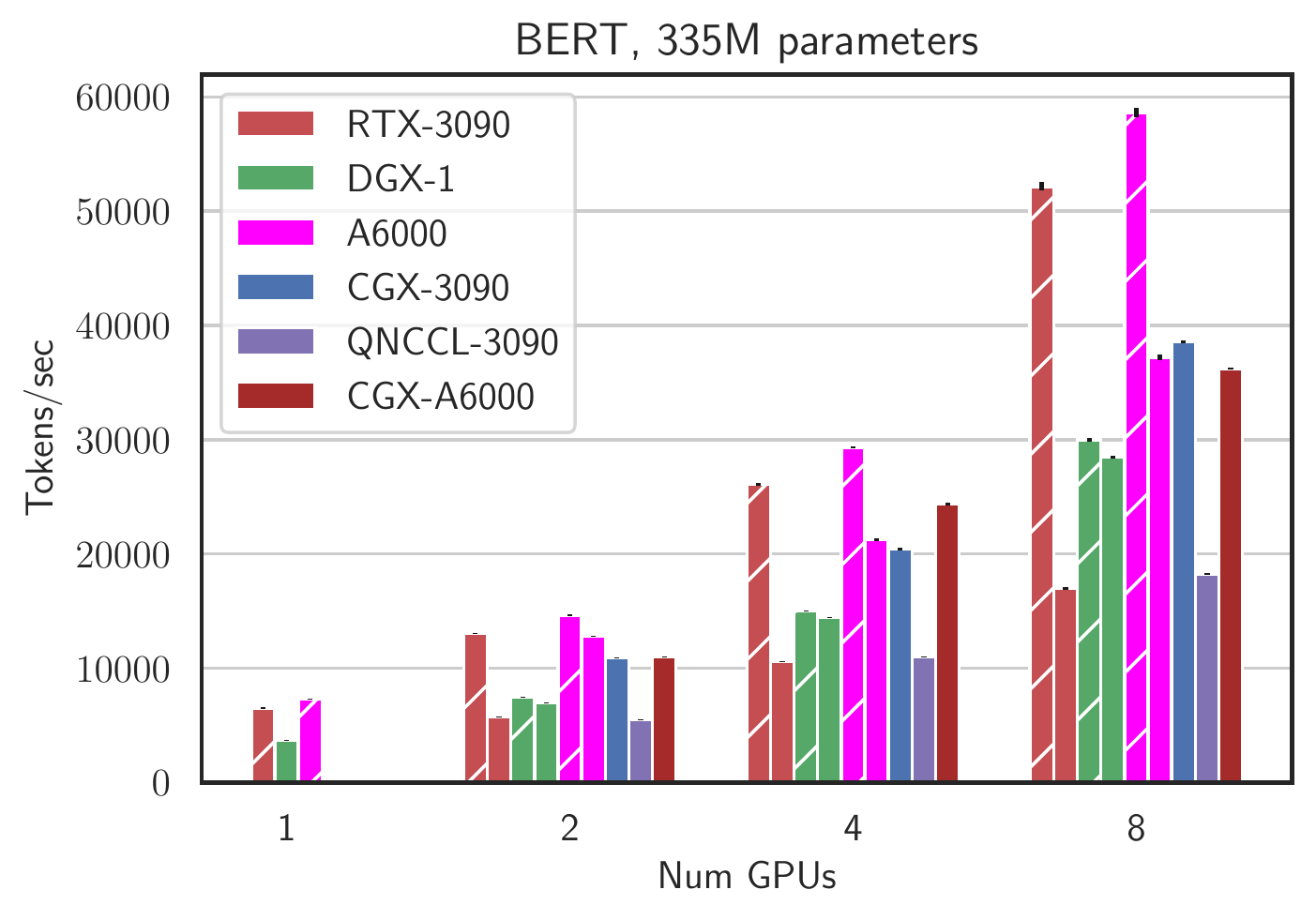}}
    \caption{{\small Throughput for ResNet50/ImageNet,  Transformer-XL (TXL) on WikiText, and BERT on SQUAD. Higher is better. Hatched bars represent ideal scaling. CGX leads to self-speedups of $>2\times$, and scalability of 80\% to 90\%. Hatched bars represent ideal scaling.}}
    \label{fig:throughput}
 \end{figure*}

\subsection{Experimental results}

\subsubsection{Accuracy Recovery.}
We first examine the model accuracies using standard hyper-parameters in end-to-end training experiments. The gradient bit-width used for these experiments is 4 bits. 
The bucket size was 1024 for CNNs, and 128 for  Transformer models, chosen empirically. As stated, we reduce small layers (biases, batch and layer normalization layers) in full precision.
The results of training on the RTX3090 machine with 8 GPUs are presented in the Table~\ref{table:val_results}, with the corresponding accuracy parameters. All CGX accuracy results are within the standard 1\% error tolerance~\cite{mattson2020mlperf}; in most cases, accuracy is within seed random variability.

Following the original recipes, ResNet50, VGG16, and the Vision Transformer (base model) were trained on ImageNet with total batch sizes 256, 256, 576 respectively. ViT was trained in mixed precision level 1 (activations at FP16, weights, and gradients in full precision). The Transformer-XL (base model) experiment was run on WikiText-103 dataset with batch size 256 and second level mixed precision (model, activations, and gradients cast FP16). The GPT-2 model was trained on WikiText-2, batch size 24, level 2 mixed precision. For question-answering we used BERT model on the SQUAD-v1 dataset with batch size 3 per GPU and FP32 training.

Unless otherwise stated, we focus on following model/task combinations: Transformer-XL on WikiText-103, ResNet50 on ImageNet, and ViT on ImageNet.
The parameters are identical to the ones provided above. All experiments were run on 8 GPUs.

% We provide detailed training details in terms of batch size and mixed precision in Appendix C. 
% \subsubsection{Performance of Individual Components.}

\begin{table}[t]
\centering
\caption{Training throughput with CGX, PowerSGD, and GRACE on single machine with 8 RTX3090 GPUs. 
(Transformer-XL/PowerSGD  did not converge, so we only provide throughput numbers.)}
\label{table:psgd_grace_comparison}

{\footnotesize
\begin{tabular}{|c|c|c|c|c|}
\hline
\centering
& ResNet50 & Transformer-XL-base & BERT\\
\hline
Baseline & 1900 & 170k & 17.5k \\
\hline
CGX  & \textbf{2900} & \textbf{260k} & \textbf{38.7k} \\
\hline
PowerSGD  & 2600  & 220k* & 38.3k\\
\hline
Grace & 1000 & 30k & 14.3k\\
\hline
\end{tabular}
}

\end{table}

%We compare the performance of various reduction schemes in Figure~\ref{fig:reductions}. 
\subsubsection{Comparison with other algorithmic approaches.}\hfill\\
\paragraph{PowerSGD Compression.}
We follow the implementation of~\cite{vogels2019powersgd}, and set the rank to 4 for CNNs and use rank 8 for Transformers, implying up to 100x compression. PowerSGD can not be used in conjunction with FP16 training, as it can lead to divergence in our experiments, so we compare at FP32. But with full-precision gradients training PowerSGD can not achieve baseline accuracy at Transformers pre-training (we tried ranks up to 32).
% Using FP32 for fairness, CGX has superior performance, by up to 2x in the case of Transformer models, and around 10-15\% speedup for CNNs. 
As Table~\ref{table:psgd_grace_comparison} shows CGX has superior performance on single node over PowerSGD in spite of lower compression.
This is because 1) higher compression shows diminishing returns, 2) CGX has lower compression overhead (Table \ref{table:compression_approaches}), and 3) CGX implements faster reductions. 

\paragraph{Sparsification.}
% TODO We can tell here that we could not make topK converge under standard parameters
We implemented the TopK \cite{dryden2016communication} algorithm as part of CGX framework. Usage of the sparcification compression there faces following issues. In order to converge under standard parameters, sparcification must be applied upon entire model, not layer-wise which is impossible due to specifics of the communication frameworks (\texttt{torch.distributed}, \texttt{Horovod}). In our experiments we did not manage to make topK with error feedback converge with similar to QSGD compression rate. Moreover, topK with higher compression rates did not show any speedup in comparison to QSGD due to compression saturation on our workstation (see Figure~\ref{fig:fake_compression}) and higher topK overhead (see Table~\ref{table:compression_approaches}, we used ~\cite{Abdelmoniem2021} topK mechanism).

\subsubsection{Comparison with other systems}\hfill\\
\paragraph{GRACE Comparison.}
We adapted our benchmarks to also compare to  GRACE~\cite{grace}, which also implements quantization and sparsity compression techniques. We used the same uniform 4-bit compression variant for frameworks, as this recovers accuracy. We used NCCL  as the communication backend for GRACE, as it provided the best performance in our setting. We found (see Table~\ref{table:psgd_grace_comparison}) that CGX outperforms GRACE by more than 3x on average. Our profiling suggests that this occurs because GRACE uses a less effective  reduction scheme (NCCL-Allgather vs. optimized Allreduce), less efficient compression (e.g., no bucketing) and transmission (even with 4 bits compression, GRACE communicates in INT8). We also tried GRACE with very high-sparsity TopK compression (0.001), and   performance did not improve significantly. This suggests that GRACE's implementation has additional bottlenecks in terms of communication latency.

\paragraph{NCCL and QNCCL Comparison.}
As shown in Figure~\ref{fig:fake_compression}, NCCL has poor scaling on commodity machines, especially from 4 to 8 GPUs, where communication cost is highest. 
CGX can give > 2x speedup relative to NCCL, reaching 80-90\% of linear scaling. This enables the consumer-grade RTX3090 GPU to match or even surpass the throughput of a DGX-1 server. We found that QNCCL partly alleviates the scaling problems of NCCL, and only improves throughput by a limited margin, as it does not benefit from the bespoke communication backend integrated in CGX. 
An orthogonal issue for QNCCL is the fact that it has higher accuracy degradation: since compression cannot be performed layer-wise (as QNNCL does not have layer information), it cannot perform layer-wise compression. We have been able to recover accuracy within 1\% with QNCCL at 4bit compression by reducing bucket size to 128 for all models, but this comes with a further performance reduction. 

% \paragraph{HiPress-CaSync Comparison.}
% The HiPress-CaSync~\cite{bai2021hipress} framework implements a communication system supporting quantization methods for a Parameter-Server setup, and Ring communication patterns. 
% We note that a complete comparison is hampered by the fact that the libraries HiPress relies on are not compatible with newer GPUs (so we could not compare on the 8x RTX 3090 machine), and that this library only supports 2-bit quantization. 
% Nevertheless, we executed a comparison on 4x EC2 p3.8xlarge instances with 4 V100 GPUs each, using the Docker image provided by HiPress, and ran image classification benchmarks, ResNet50 and VGG19 as in the original paper. The results for 2-bit compression showed that CGX and HiPress-CaSync have similar performance on the smaller ResNet50 model, and that CGX outperforms HiPress by approximately 5\% when training VGG19. 

\paragraph{Bagua and HiPress Comparison.}
Bagua~\cite{gan2021bagua} and HiPress-CaSync~\cite{bai2021hipress} are distributed training frameworks, which also support some generic forms of gradient quantization.
In multinode experiments on 4x EC2 p3.8xlarge instances with 4 V100 GPUs each, we observed that Bagua and HiPress have similar performance to CGX on the smaller ResNet50 model, and that they are up to 10\% slower on the larger VGG19 model. This is since all frameworks use the same NCCL backend for inter-node communication, but CGX uses a faster pattern (SRA vs Ring or Tree for NCCL) than HiPress and has better compression rate than Bagua (which only supports 8 bit quantization). Moreover, HiPress only supports 2 bit quantization, e.g. \cite{wen2017terngrad,strom2015scalable}), which does not converge under standard parameters for Transformed-based models.

\begin{figure}
    \centering
    \setkeys{Gin}{width=0.49\linewidth}
    \subfloat[ResNet50]{\includegraphics{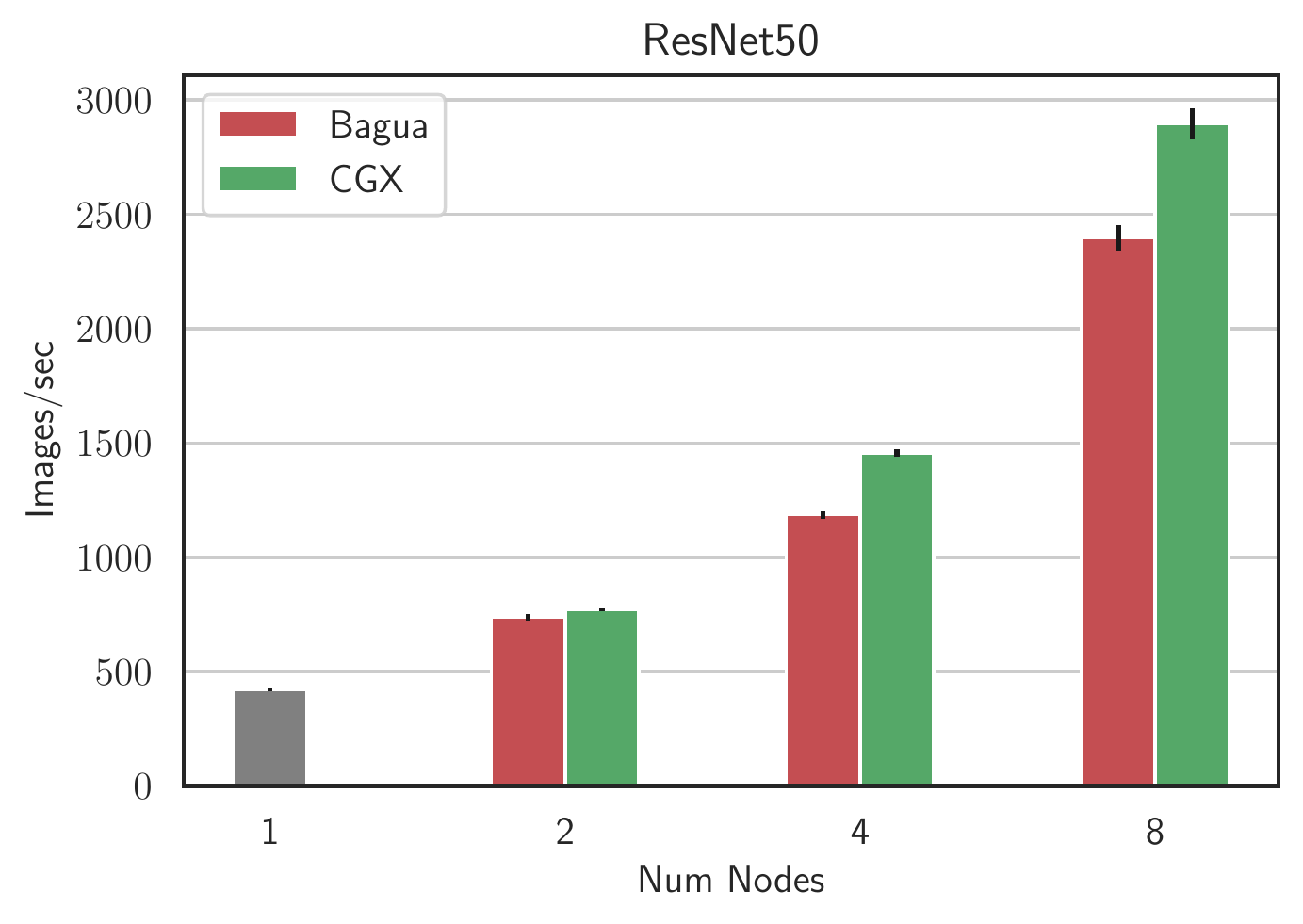}}\hfil
    \subfloat[VGG19]{\includegraphics{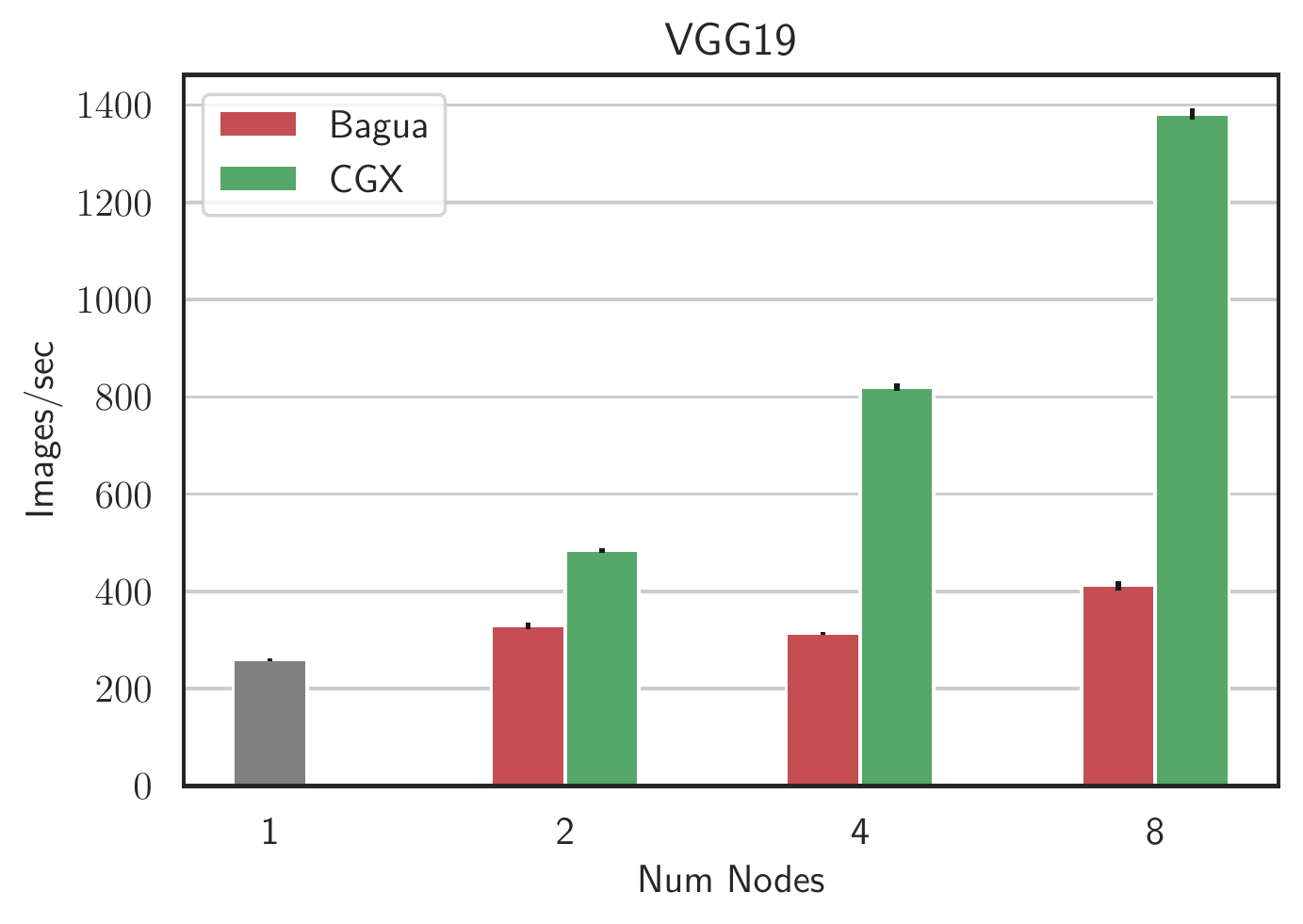}}
    \caption{Scaling throughput in multi-node environment for image classification tasks. Bagua vs CGX.}
    \label{fig:bagua}
\end{figure}

HiPress unfortunately does not support the newer commodity RTX-3090 GPUs, so 
we could only compare with Bagua on the Genesis Cloud 8xRTX3090 instance. 
The results of the comparison are presented in Figure~\ref{fig:bagua}, showing that CGX provides clearly superior performance, especially for the VGG19 model. 

% In order to compare CGX with Bagua allreduce we ran experiments on Genesis 8xRTX3090 instance. Bagua with gradient compression performed by up to 15\% slower than Bagua without compression on a single node so we used uncompressed Bagua communication. The results of the comparison are presented in Figure~\ref{fig:bagua}.

\subsubsection{Comparison with Hardware Bandwidth Overprovisioning.}
We now turn to Figure~\ref{fig:throughput} where we first observe that, although in terms of single-GPU performance the RTX3090 is comparable to the V100/DGX-1, it has poor multi-GPU scaling for large models when using the standard NCCL setup ($< 50\%$ of linear scaling). The older 2080 GPUs have lower throughput both due to both lower memory, limiting maximum batch size, as well as lower computational power (Fig.~\ref{fig:throughput-rn50}). 
Thus, we mainly focus on 3090 GPUs. 

% has the worst performance as expected, CGX-RTX2080 considerably improves NCCL but even linear scaling can not reach V100 performance. RTX2080 does not have enough GPU RAM for large models experiments and reducing batch size only enlarges the gap between the machines performance results. Therefore, in further experiments we consider only RTX3090 machine.

%In case of Vision Transformer experiment, RTX3090 machine surpasses DGX-1 as the training is not bandwidth-bottleneck due to high costs of computation. Nevertheless, CGX allows us to get additional 10\% speedup approaching the theoretically maximal throughput.

% \paragraph{Power Draw.}
% The RTX-3090 GPU has higher maximal power draw than the V100 (see Table~\ref{table:GPU_char});  nevertheless, in our experiments we observed similar power consumption via \texttt{nvidia-smi}, given the same workload. Since currently no precise global measurement tools are available, we leave an exact comparison for future work.

If we compare the maximum achievable performance (ideal scaling), CGX achieves similar results to the bandwidth overprovisioning approach, on both the DGX and the A6000 machines. In other words, CGX allows us to get bandwidth-overprovisioning performance via a ``middleware'' approach, achieving our stated goals. The remaining percentage gaps from perfectly linear scaling are because of 1) latency costs, 2) inefficiencies in our implementation, and 3) remaining communication costs, especially in early layers, which cannot be overlapped with computation. To measure this, we artificially removed the bandwidth bottleneck, by sending only a small number of elements per layer. 
The results in Table~\ref{table:fake_latency} show that CGX is close to ideal bandwidth reduction.  

\begin{table}[h]
\centering
\caption{Ideal performance (\% of linear scaling) achievable via bandwidth-overprovisioning for different workloads, relative to CGX.}
\label{table:fake_latency}
{\footnotesize
\begin{tabular}{|l|l|l|l|l|l|}
\hline
 & ResNet50    &  VGG16  & TXL  & BERT & ViT  \\
\hline
Ideal Perf. & 92  \%  & 91 \%  &  95 \% & 88 \% & 95 \% \\
\hline
CGX Perf. & 90\% & 84 \% & 87 \% & 75 \% & 93 \% \\
\hline
\end{tabular}
}
\end{table}

\begin{figure}
    \centering
    \includegraphics[width=0.4\textwidth]{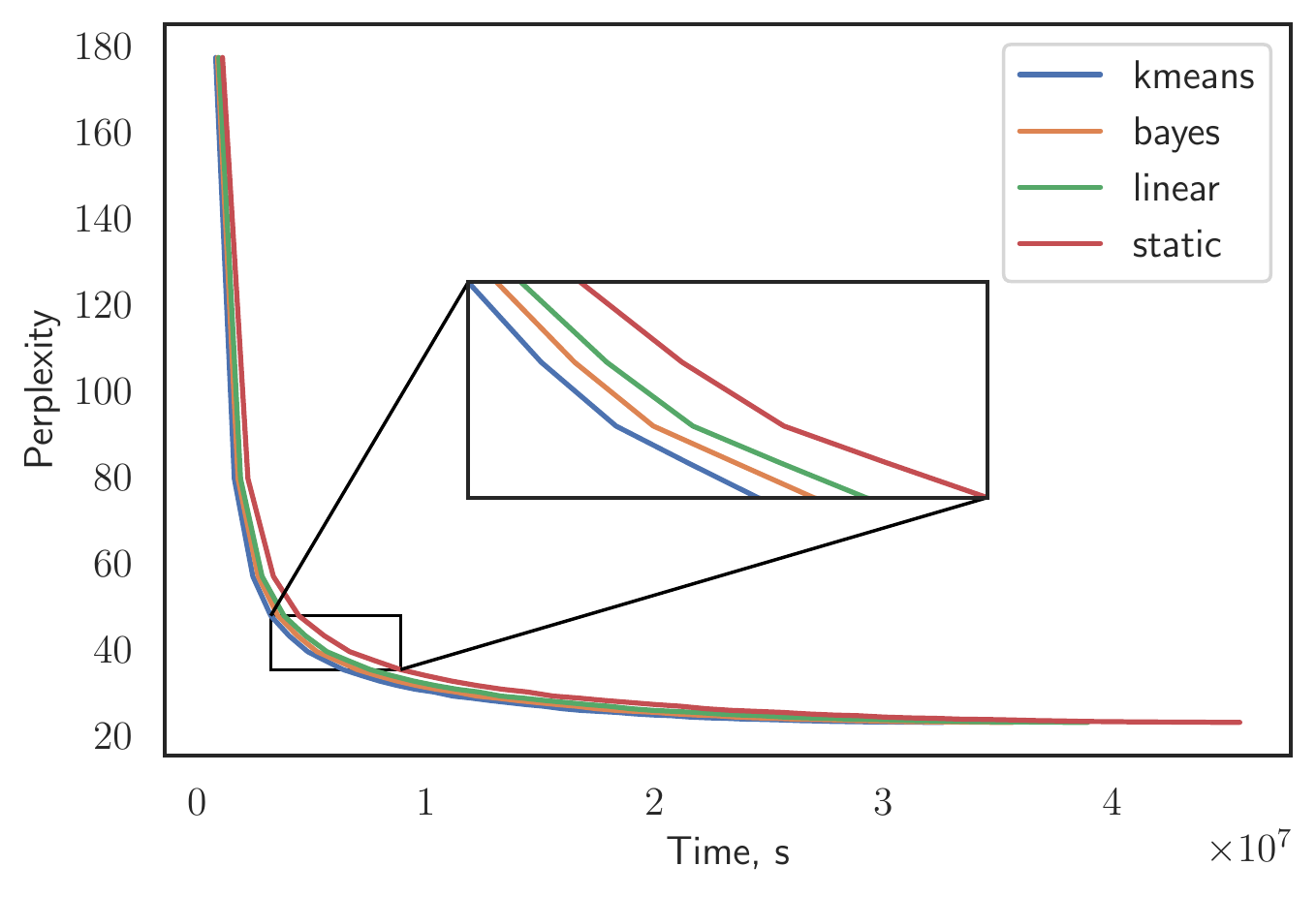}
    \caption{Transformer-XL training with adaptive schemes.}
    \label{fig:adapt_ppl_vs_time}
\end{figure}

\begin{figure}
    \centering
    \setkeys{Gin}{width=0.49\linewidth}
    \subfloat[Compression error.\label{fig:adapt_compression_error}]{\includegraphics{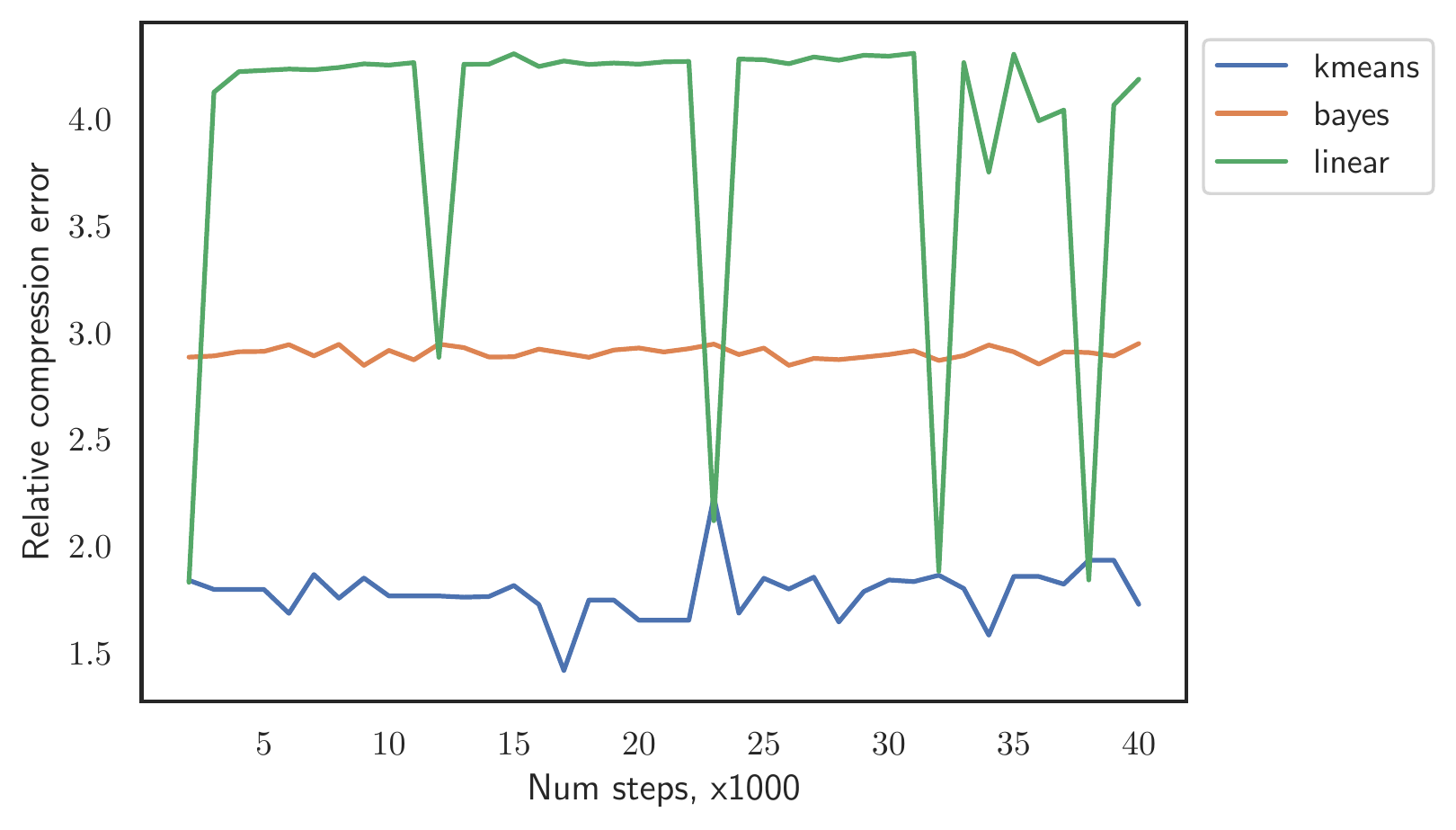}}\hfil
    \subfloat[Compression ratio.\label{fig:adapt_compression_ratio}]{\includegraphics{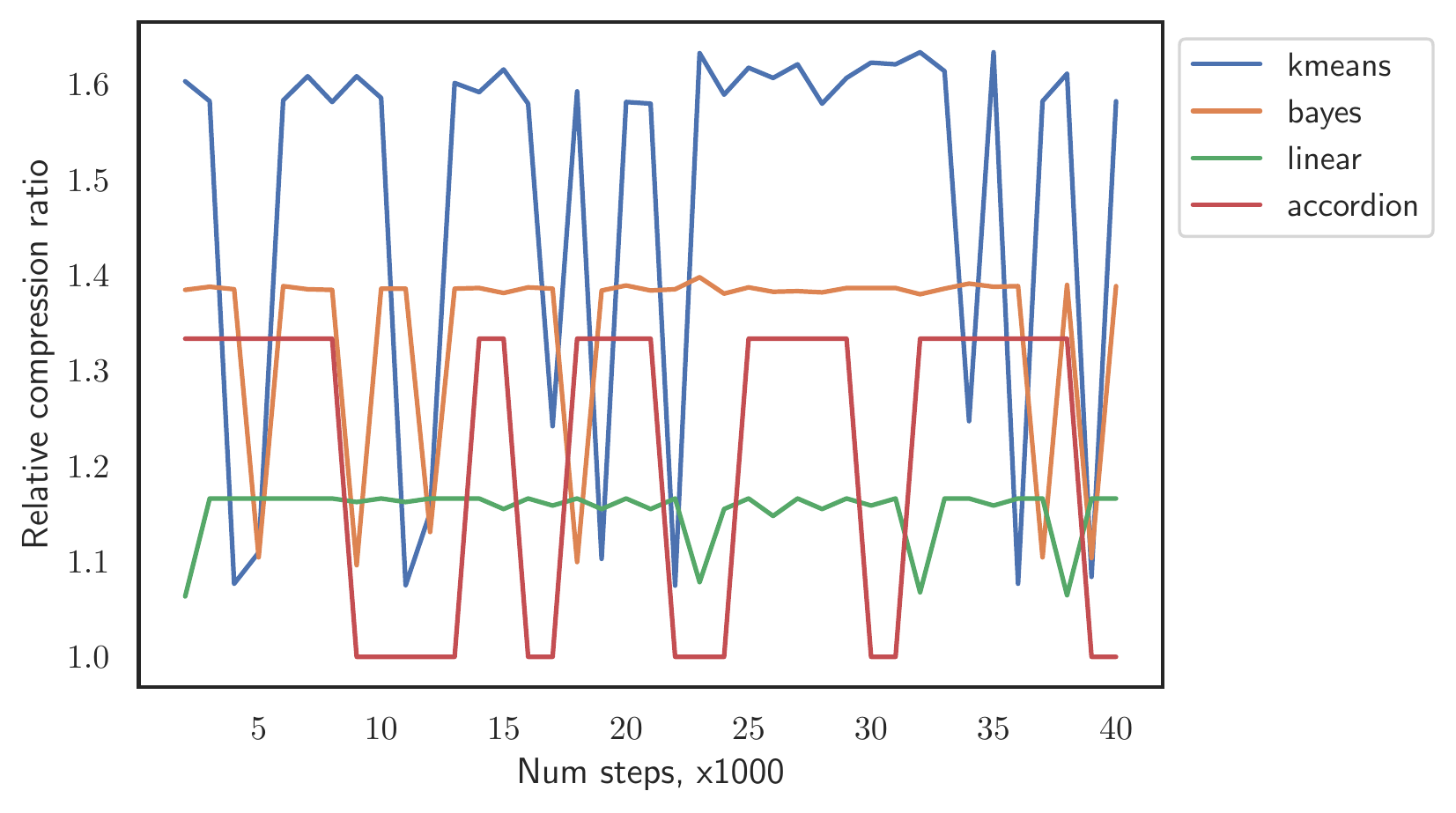}}
    \caption{Comparison of adaptive compression approaches. Error and size compression are shown relative to uniform static assignment of compression parameters to 4 bits.}
\end{figure}

% \begin{figure}
%      \centering
%      \includegraphics[width=0.4\textwidth]{images/Adapt_compression_error.pdf}
%      \caption{Compression error throughout the training for different adaptive algorithms (relative to static assignment).}
%      \label{fig:adapt_compression_error}
% \end{figure}

%  \begin{figure}
%      \centering
%      \includegraphics[width=0.4\textwidth]{images/Adapt_compression_ratio.pdf}
%      \caption{Compression ratio throughout the training for different adaptive algorithms(relative to static assignment).}
%      \label{fig:adapt_compression_error}
% \end{figure}

\begin{table}[h]
\centering
{
\caption{Comparison of adaptive methods. Speedups and compression rates are relative to static bits-width assignment (4 bits). Experiments are run with Transformer-XL base model on 8 RTX3090 GPUs (single node) and 4 nodes with 4xRTX3090 GPUs each (multi-node). Accordion is applied to QSGD with 3 and 4 as compression bounds.}
\label{table:adaptive_methods_comparison}
{\footnotesize
\begin{tabular}{|c|c|c|c|}
\hline
\centering
& Compression  & Speedup 1-Node & Speedup Multi-Node \\
\hline
KMEANS & \textbf{1.47} & \textbf{5}\% & \textbf{40}\% \\
\hline
Bayes & 1.34 & 3\% & 30\% \\
\hline
Linear & 1.15 & 2\% & 13\% \\
\hline
Accordion & 1.21 & 3\% & 15\% \\
\hline 
\end{tabular}
}
}
\end{table}

\subsection{Layer-wise Adaptive Compression} 
\label{sec:adaptive}

So far, we have provided results for our version of 4bit quantization, which always recovers accuracy. 
We now examine additional performance savings due to adaptive compression. 
Across all models, the automated procedure in Section~\ref{sec:layer_filters} identifies large layers with low-performance sensitivity (e.g. fully-connected or embedding layers) for lower bit-widths, and has similar total compression error to uniform compression.
We illustrate this on Transformer-XL, the model with the most non-uniform layer sizes. We conducted single-node experiments on an 8xRTX3090 machine, and multi-node on four 4xRTX3090 machines. As before, the baseline is 4-bits static compression, which was shown to recover full accuracy.
Figure~\ref{fig:adapt_ppl_vs_time} represents perplexity against time for different selection mechanisms. Figures~\ref{fig:adapt_compression_error} and~\ref{fig:adapt_compression_ratio} represent compression error and compression ratio relative to static assignment.  Table~\ref{table:adaptive_methods_comparison} shows that Bayesian optimization shows stable compression error, and good \emph{average} compression.  Yet, the kmeans-based method shows the lowest quantization error, best average compression, and highest speedup, as it tends to compress large  layers more. 
Specifically, this can lead to additional improvements in the order of 5\% on a single node and up to 40\% in multinode setting, without accuracy loss. This approach can still be improved by taking into account \emph{runtime speedups} instead of absolute compression. 

Among existing adaptive schemes,  AdaComp~\cite{Chen2018} and Accordion~\cite{Agarwal2021} are the only ones which can be adapted to our setting. AdaComp suggests an adaptive scheme for sparsification, with possible further quantization of communicated elements. Accordion adapts gradient compression parameters based on identifying critical learning regimes.
%Both of these adaptive systems output only global compression parameters, i.e. they are applied to all layers uniformly.

For comparison, we execute the Transformer-XL model on a language modelling (LM) task.  We first applied AdaComp only for sparsification: however, unfortunately the compression assignment provided by AdaComp did not converge to reasonable accuracy on this task.

Second, we adapted Accordion to our framework with QSGD compression, using Accordion to choose bit-width parameters based on its critical regimes detection approach. We used Accordion with hyperparameter $\eta=0.5$, as suggested by the authors, and updated the compression parameter every 1k steps of training. As the lower and higher compression levels, we checked  (2, 4) and (3, 4). The first pair resulted in significantly lower final accuracy relative to the baseline. The second pair (3,4) recovered the final accuracy, but the compression ratio was inferior to all the other adaptive schemes we investigated, and considerably below our proposed clustering scheme. Please see Figure~\ref{fig:adapt_compression_ratio} and Table~\ref{table:adaptive_methods_comparison} for an illustration. The table represents the speedups of different adaptive methods relative to the regular static compression.
For instance, our adaptive scheme resulted in 17\% additional multi-node speedup compared to Accordion.
% TODO Add Accordion results to the figure.
% Therefore, 
 
% This may appear small; yet, as discussed above, this converges to the point at which bandwidth is no longer a bottleneck. 

% \paragraph{Heterogeneous compression.} 
% CGX also allows customized compression, e.g. sparsification on specific layers and quantization on other layers: this could be a good option for Transformer models, whose embedding layers are naturally sparse. 
% We did experiment with this approach, specifically applying TopK-SGD~\cite{dryden2016communication} with error feedback, transmitting only 1\% of parameters. 
% Yet, we only obtain a modest additional 3\% speedup over quantization. This is justified due to the additional cost of TopK compression, but also due to the fact that our system is already close to ideal in terms of bandwidth savings when using quantization.

\subsection{Practical Implications}

\paragraph{Multi-node experiments.}
Next, we examine performance on multi-node training in the cloud. We used 4 4xRTX3090 Genesis instances with 10GBps intra-node bandwidth and 5 GBps inter-node bandwidth. Table~\ref{table:multi_node_results} shows that CGX provides up to 10x speedup over the uncompressed baseline.

\begin{table}[h]
\centering
\caption{Items per second when training with the NCCL and CGX optimizations, respectively, on 4 machines with 4 RTX3090 GPUs each.}
\label{table:multi_node_results}

{\footnotesize
\begin{tabular}{|c|c|c|c|c|}
\hline
\centering
& ResNet50 & ViT-base & Transformer-XL-base & BERT\\
\hline
Baseline & 564 & 34 & 32k  & 1.4k \\
\hline
CGX  & 2.3k & 235  & 85k & 12k\\
\hline
\end{tabular}
}

\end{table}

\paragraph{Implications for Cloud Training.}
Several cloud services provide servers with commodity GPUs~\cite{GenesisCloud, LambdaCloud, LeaderGPU}. 
We therefore compare a standard AWS EC2 4xV100 GPU instance (p3.8xlarge)  instance with a 4xRTX 3090 Genesis Cloud instance~\cite{GenesisCloud}. We execute the same training benchmark, with and without CGX. 
The numbers in Table~\ref{table:genesis_cloud} show that CGX allows us to obtain almost \emph{twice} higher throughput (training tokens/second) per dollar on the more affordable Genesis instance, for a standard  language modelling task (SQuAD) task using an industry-standard BERT model.

\begin{table}[h]

{\footnotesize
\centering
\caption{Comparison of training performance for different cloud services (AWS and Genesis) with and without CGX. 
The training task is BERT-QA and achieves full accuracy.}
\label{table:genesis_cloud}
\begin{tabular}{|p{20mm}|p{15mm}|p{15mm}|p{17mm}|}
\hline
Instance & Throughput (1K tok./sec)    &  Price per hour (\$)  & Tokens/second per \$  \\
% \hline
% Genesis  NCCL & 105.4 & 13.6 & 7.75 \\
% \hline
% AWS V100 NCCL & 258.8 & 24.5 & 10.6 \\
% \hline
% AWS A100 NCCL & 449.8 & 32.7 & 13.7 \\
% \hline
% Genesis \textbf{CGX} & 173 & 13.6 & 12.72 \\
% \hline

\hline
Genesis + NCCL & 4737 & 6.8 & 696 \\
\hline
AWS + NCCL & \textbf{14407} & 12.2 & 1181 \\
\hline
Genesis + {CGX} & \textbf{14171} & 6.8 & \textbf{2083} \\
\hline
\end{tabular}
}
\end{table}

\vspace{-0.5em}
\section{Conclusions}
We proposed an algorithms \& systems approach to remove the bandwidth bottlenecks from DNN training, supplanting the need for dedicated hardware support, and significantly improving performance in both single node (commodity) settings and, more generally, in multi-node cloud settings.  
Future work may extend our results to model-parallel or hybrid synchronization setups, e.g.~\cite{zhou2020petrel, li2021sync}; moreover, the idea of adaptive layer-wise compression should be extensible to other compression methods, for instance to choose ranks accurately for gradient decomposition methods, or layer-wise sparsities based on actual transmission speedups. 

\begin{acks}
The authors sincerely thank Nikoli Dryden, Tal Ben-Nun, Torsten Hoefler, Bapi Chatterjee, Tim Harris and Sylvain Jeaugey for useful discussions throughout the development of this project. This project has received funding from the European Research Council (ERC) under the European Union's Horizon 2020 research and innovation programme (grant agreement No 805223 ScaleML).
\end{acks}

% \printbibliography
\bibliographystyle{ACM-Reference-Format}
\bibliography{references}

\end{document}